\documentclass[aoas,preprint]{imsart}

\usepackage[toc,page]{appendix}
\RequirePackage[OT1]{fontenc}
\RequirePackage{amsthm,amsmath}
\RequirePackage{natbib}
\RequirePackage[colorlinks,citecolor=blue,urlcolor=blue]{hyperref}

\startlocaldefs
\numberwithin{equation}{section}
\theoremstyle{plain}

\endlocaldefs

\usepackage{caption}
\usepackage{multirow}
\usepackage{comment}
\usepackage{amsmath}
\usepackage{float}
\usepackage{tabularx}
\usepackage{color}
\usepackage{graphicx}
\usepackage{ulem}
\usepackage{enumerate}
\usepackage{textcomp}
\usepackage{gensymb}
\usepackage{multicol}
\usepackage{caption}
\usepackage{subcaption}

\newcommand{\logrhk}{\log R'_{\text{HK}}}
\newcommand{\bis}{\text{BIS}}
\newcommand{\kernel}{K}

\newcommand{\gp}{X}
\newcommand{\extragp}{Z}

\newcommand{\indnum}{L}

\newcommand{\edits}{\color{black}}

\defcitealias{rajpaul2015gaussian}{R15}

\begin{document}

\begin{frontmatter}
\title{Improving Exoplanet Detection Power: Multivariate Gaussian Process Models for Stellar Activity}
\runtitle{Improving Exoplanet Detection Power}
\thankstext{T1}{%
This material is based upon work supported by the National Science Foundation under Grants No. DMS-1127914 (to the Statistical and Applied Mathematical Sciences Institute), AST-1616086 (for E.~Ford), AST-1312903 (for T.~Loredo), and DMS-1622403, SES-1521855, and ACI-1550225 (for R.~Wolpert). E.~Ford acknowledges support by the Institute for CyberScience and the Center for Exoplanets and Habitable Worlds, which is supported by Pennsylvania State University, the Eberly College of Science, and the Pennsylvania Space Grant Consortium. E.~Ford also acknowledges support from NASA Exoplanets Research Program grant \#NNX15AE21G and supporting collaborations within NASA's Nexus for Exoplanet System Science (NExSS). T.~Loredo acknowledges support from NASA Astrophysics Data Analysis Program grant \#NNX16AL02G. X. Dumusque acknowledges the Society in Science's Branco Weiss Fellowship for its financial support.}

\begin{aug}
\author{\fnms{David E.} \snm{Jones}\thanksref{tamu}\ead[label=e1]{david.jones@tamu.edu}},
\author{\fnms{David C.} \snm{Stenning}\thanksref{sfu}}
\author{\fnms{Eric B.} \snm{Ford}\thanksref{psu}},
\author{\fnms{Robert L.} \snm{Wolpert}\thanksref{duke}},
\author{\fnms{Thomas J.} \snm{Loredo}\thanksref{ccaps}}, \author{\fnms{Christian} \snm{Gilbertson}\thanksref{psu}},
\and
\author{\fnms{Xavier} \snm{Dumusque}\thanksref{ug}}

\runauthor{Jones et al.}

\affiliation{Texas A\&M University\thanksmark[1]{tamu}, Simon Fraser University\thanksmark[2]{sfu}, Penn State University\thanksmark[3]{psu}, Duke University\thanksmark[4]{duke}, Cornell Center for Astrophysics and Planetary Science\thanksmark[5]{ccaps}, and Observatoire Astronomique de l'Universit\'e de Gen\`eve\thanksmark[6]{ug}}

\end{aug}

\begin{abstract}
The radial velocity method is one of the most successful techniques for detecting exoplanets. It works by detecting the velocity of a host star induced by the gravitational effect of an orbiting planet, specifically the velocity along our line of sight, which is called the {\it radial velocity} of the star.  Low-mass planets typically cause their host star to move with radial velocities of 1 m/s or less. By analyzing  a time series of stellar spectra from a host star, modern astronomical instruments can in theory detect such planets. 
However, in practice, intrinsic stellar variability (e.g., star spots, convective motion, pulsations) affects the spectra and often mimics a radial velocity signal. This signal contamination makes it difficult to reliably detect low-mass planets. A principled approach to recovering planet radial velocity signals in the presence of stellar activity was proposed by \citet{rajpaul2015gaussian}. It uses a multivariate Gaussian process model to jointly capture time series of the apparent radial velocity and multiple indicators of stellar activity. We build on this work in two ways: (i) we propose using dimension reduction techniques to construct new high-information stellar activity indicators; 
and (ii) we extend the \citet{rajpaul2015gaussian} model to a larger class of models and use a power-based model comparison procedure to select the best model. 
Despite significant interest in exoplanets, previous efforts have not performed large-scale stellar activity model selection or attempted to evaluate models based on planet detection power.  
In the case of main sequence G2V stars, we find that our method substantially improves planet detection power compared to previous state-of-the-art approaches. 
\end{abstract}



\end{frontmatter}

\section{Introduction}

\subsection{Motivation}
\label{sec:intro}

In this paper, we present a statistical framework to improve the sensitivity of astronomical surveys for detecting exoplanets, i.e., planets orbiting stars other than the Sun. As a planet orbits a star, the planet's gravitational force causes the star to orbit around the center of mass of the system. Consequently, the starlight appears to be alternately shifted to longer (redder) and shorter (bluer) wavelengths, as the star moves away from and towards the observer, respectively. That is, the observed starlight contains a periodic signal due to the Doppler effect resulting from the star's motion.  

\begin{figure}[t]
\centering
\begin{subfigure}{.55\textwidth}
  \centering
  \includegraphics[width=1\linewidth]{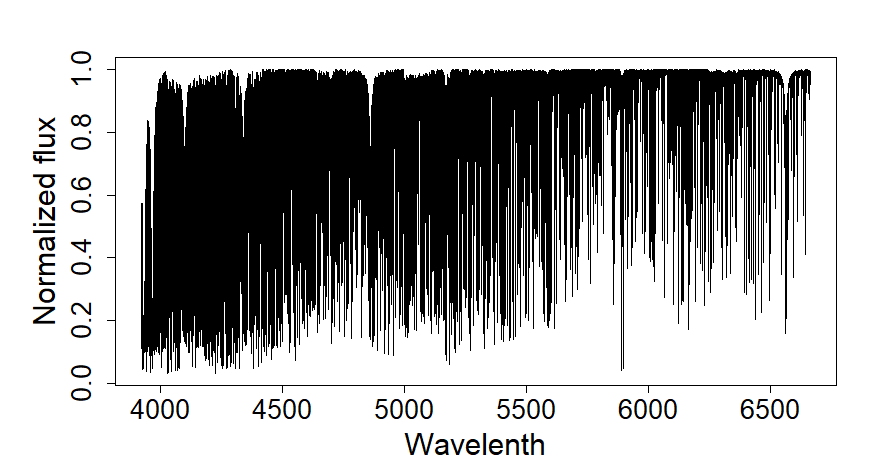}
\end{subfigure}%
\begin{subfigure}{.45\textwidth}
  \centering
  \includegraphics[width=1\linewidth,trim=0mm 10mm 0mm 0mm,clip]{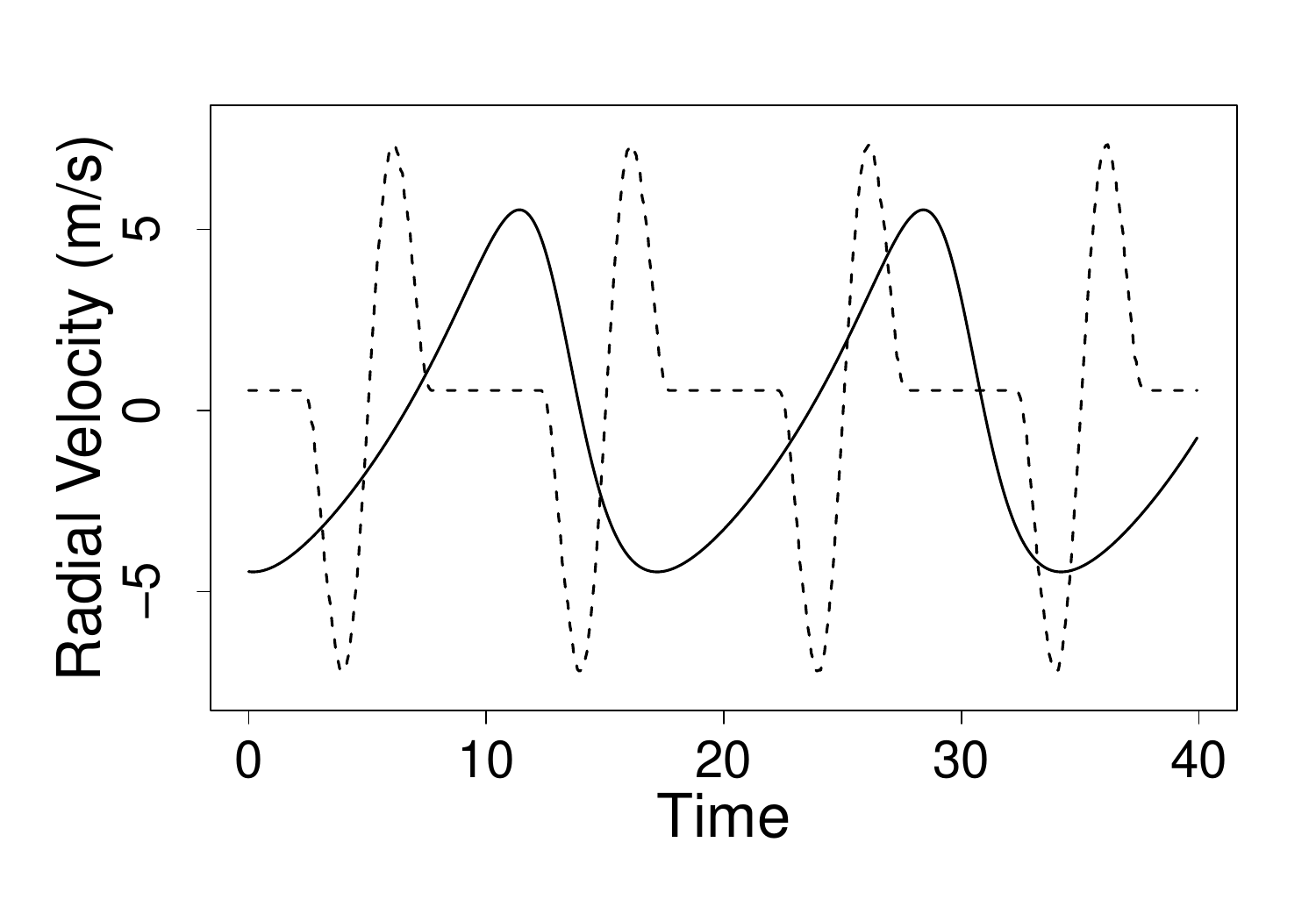}
\end{subfigure}
\caption{Example stellar spectrum (left) and an illustration of the scientific model for a RV signal due to a planet (right, solid line) and the RV corruption due to a spot (right, dashed line). 
\label{fig:spectrum}}
\end{figure}

In radial velocity exoplanet surveys, the raw data are a time series of high-dimensional spectra, and it is from these that the magnitude of the Doppler effect is measured, as we now explain. A spectrum is the intensity of starlight as a function of wavelength, and an example spectrum is shown in the left panel of Figure \ref{fig:spectrum}. The spectrum contains many tens of thousands of spectral ``lines'' (i.e., dips in the stellar spectrum) because the star's atmosphere absorbs  light at specific wavelengths. The Doppler effect is detected by measuring how the observed wavelengths of the spectral lines are shifted over time. In particular, if $f$ denotes the unperturbed stellar spectrum, and $z(t)$ denotes the Doppler shift at time $t$, then the observed spectrum at time $t$ and wavelength $\lambda$ is given by 
\begin{align}
    f_t(\lambda) = 
    f\left(\frac{\lambda}{1+z(t)}\right) + \epsilon_{t\lambda},\label{eqn:redshift}
\end{align}
where $\epsilon_{t\lambda}$ denotes noise. 
While the Doppler shifts $z(t)$ due to a planet are very small, modern astronomical instrumentation can detect them by jointly analyzing many spectral lines, e.g., \citet{butler1996attaining,baranne1996elodie,phase2003setting}.  

In practice, the computed Doppler shifts are typically converted to a more physically interpretable quantity, namely the velocity of the star projected onto the observer's line of sight, which is known as the {\it radial velocity} (RV) of the star. The radial velocity is denoted $v(t)$ and is given by $v(t)=cz(t)$, where $c$ is the speed of light. Currently, radial velocities can be determined down to $\sim$1 m/s \citep{fischer2016state}, and upcoming instruments will likely be even more sensitive. 
To hunt for planets, astronomers choose candidate stars and for each collect a time series of RV observations spanning many days up to years, e.g., \citet{mayor2011harps,pepe2011harps,fischer2013twenty,butler2017lces}. The shape of the RV signal expected due to a planet is well understood based on Newton's laws of motion and gravity. The right panel of Figure \ref{fig:spectrum} shows an example RV signal produced by the scientific model for a large planet with a 17 day orbital period (solid line). 
Hunting for periodic RV signals such as this 
is known as the {\it radial velocity method} and is one of the most successful methods for detecting exoplanets. 
However, astronomers would like to reach below the current detection threshold of 1 m/s because many low-mass planets  induce RV signal  amplitudes smaller than this threshold, e.g., Earth induces an RV signal of about 9 cm/s. 


A key challenge for the radial velocity method in practice, and the focus of this paper, is that planet RV signals can be mimicked by intrinsic stellar variability, i.e., phenomena occurring in the outer part of the observed star. 
For instance, stellar magnetic fields can lead to groups of small, dark and relatively cool {\it star spots}, often surrounded by brighter regions known as {\it faculae}.
As a monitored star rotates, any spots (or faculae) will move across the visible stellar disk and lead to a spurious signal in the observed RV time series. 
Stellar rotation periods are often similar to plausible planetary orbital periods and therefore distinguishing spots and faculae from planets can be particularly problematic, e.g., the signal discovered by  \citet{dumusque2012earth} was later found to be spurious by \citet{rajpaul2015gaussian} and \citet{rajpaul2015ghost}. 
An example of the RV corruption due to a single, non-evolving star spot is shown by the dashed line in the right panel of Figure \ref{fig:spectrum}. This dashed line and the solid line of the example planet RV signal are easily distinguished, but in practice separation is much more difficult due to sparse observations, noise, unknown periods, the temporal evolution of active regions, and the potential for multiple active regions at once.

Despite these difficulties, it is often in principle possible to distinguish stellar activity  from a true planet signal by using more of the raw data, rather  than only computing the RV time series. 
Specifically, we can exploit the fact that 
 a true Doppler shift affects the entire spectrum in the same way (i.e., {\it all} the spectral lines are shifted, see (\ref{eqn:redshift})), whereas  the effect of intrinsic stellar activity is different for each spectral line. 
 However, progress towards developing statistical frameworks that make use of this fact has so far been limited.  Recently, \citet{davis2017insights} used principal component analysis (PCA) to construct multiple univariate measures of stellar activity, known as stellar activity indicators, based on low-noise spectra. They showed that the evolution of these indicators over time is  different for stellar activity and Doppler shifts. 
On the other hand, they did not provide a full statistical method for making use of their stellar activity indicators to detect planets. Meanwhile, \citet{rajpaul2015gaussian} proposed a detection approach based on a sophisticated Gaussian process model that jointly captures an RV  time series and some  conventional 
stellar activity indicators. 
However, 
the impact of their modeling framework  on planet detection power 
is unclear. 


This paper unifies and extends the work of \citet{davis2017insights} and \citet{rajpaul2015gaussian} to propose a powerful framework {\edits for selecting stellar activity models and developing} new high-information stellar activity indicators. 
{\edits Currently, most exoplanet detection methods either do not use statistical models for stellar activity at all, or use a single convenient or physically motivated model, without any  empirical or theoretical statistical validation of the detection performance. To address this limitation,} {\edits this paper provides a} framework {\edits which can} evaluate and compare a large number of stellar activity models in terms of statistical power for detecting exoplanets, {\edits thereby facilitating automated and substantially improved model selection}. 
This automated approach also makes it easy to use novel stellar activity indicators, such as those of \citet{davis2017insights}, whereas previous approaches have generally been restricted to specific indicators.   
Our second contribution is to propose a modified version of the  \citet{davis2017insights} indicator construction technique. Specifically, we introduce a simple pre-processing step to separate the indicators from RV signals (real or corrupted), and thereby preserve interpretability and facilitate modeling. 
For both these new activity indicators and some conventional ones, we show that applying our model selection procedure {\edits identifies models that have high power for detecting small exoplanets in the presence of both  small and large spots, whereas existing approaches either do little to address stellar activity, or have substantially lower power in the case of large spots.} 


We recognize that exoplanet detection is a developing area of research and 
our work does not address all the scientific and statistical challenges posed by the analysis of exoplanet survey data. 
In this paper we restrict our focus to spot related activity on main sequence G2V (MS-G2V) stars, i.e., stars similar to the Sun. An initial focus on spots is reasonable because, although other forms of stellar variability\footnote{We will mainly use the term {\it stellar activity} because in astronomy this tends to refer to magnetically active regions (e.g., spots), whereas {\it stellar variability} is more general. For further details on the nature of stellar variability the reader is referred to the relevant astronomy literature, e.g., \citet{jenkins2013observing}; 
\citet{dumusque2014soap}; \citet{borgniet2015using}; \citet{haywood2016sun}.} also affect the stellar spectrum, they typically cause spurious RV signals that change on relatively short timescales, e.g., see \citet{del2004dynamics,del2004solar,arentoft2008multisite}. Secondly, to carefully study exoplanet detection power for  MS-G2V stars with spot activity we need a ground truth dataset which unfortunately is not available due to the inherent challenges in astronomy, i.e., there is no way to know for sure what type of activity a distant star has or whether a planet is orbiting it.  
To overcome this challenge we construct a realistic dataset of 1800 MS-G2V stars with various stellar spot configurations by synthesizing spectra of the Sun using the Spot Oscillation and Planet (SOAP) 2.0 software, which was developed by \citet{boisse2012soap} and \citet{dumusque2014soap}. 
Use of SOAP 2.0 data is a well established approach for developing methods to mitigate stellar activity,
e.g., \citet{rajpaul2015gaussian}; \citet{dumusque2016radial}; \citet{davis2017insights}, though previous SOAP 2.0 based investigations have not specifically studied detection power. 
{\edits  The recent work of \citet{damasso2019biases} did consider detection performance (and various biases) associated with different activity models, using simulations of stellar activity from a quasi-periodic Gaussian process. However, our SOAP 2.0 simulations are in some ways more realistic, being based on real spectra, and we compare a much larger number of models, which are also more sophisticated, e.g., we consider the  \citet{rajpaul2015gaussian} model, and extend it, whereas \citet{damasso2019biases} do not.}

This article is organized as follows. Section \ref{sec:pseudocode} overviews the paper and provides pseudocode for the key algorithms. Section \ref{sec:data} introduces the raw input data and the SOAP 2.0 package. Section \ref{sec:gpca} presents our approach for defining high-information stellar activity indicators.   Section \ref{sec:methodology} presents our general class of statistical models for jointly capturing stellar activity indicators and the RV time series. Section \ref{sec:model_selection_procedure} then introduces our model selection procedure. Section \ref{sec:selection} applies our model selection procedure and compares the planet detection power under the top performing models to that obtained using the \citet{rajpaul2015gaussian} model. Section \ref{sec:discussion} discusses the results, implications for future RV planet surveys, and areas for further research. Our code and dataset are available on GitHub at \href{https://github.com/djones2013/improving-planet-detection-power}{https://github.com/djones2013/improving-planet-detection-power}.

\section{Overview and key algorithms}
\label{sec:pseudocode}

Our paper contains five main components:\\ 
\indent\indent {\bf Part 1.} Stellar activity data and pre-processing \\
\indent\indent{\bf Part 2.} Defining new stellar activity indicators \\
\indent\indent{\bf Part 3.} Model class for stellar activity indicators \\
\indent\indent{\bf Part 4.} Ideal power-based model  selection \\
\indent\indent{\bf Part 5.} Practical power-based model selection\\
We now overview each of these parts and explain how they fit together. 
Details will be given in Sections \ref{sec:data}--\ref{sec:selection}. 

\subsection{Stellar activity dataset synthesis and pre-processing} 
\label{sec:data_overview}

 The raw data to be used in simulating spectra consists of highly precise Solar spectra from both active and quiet  
 subregions of the Sun,  a main sequence G2V star, and was collected by  \citet{wallace1998atlas} and \citet{wallace2005atlas}. The raw data can be considered ground-truth stellar activity because the effects of planets in the Solar System are well known and were accounted for in the data collection process.  

With this raw data as input, we apply the Spot Oscillation and Planet (SOAP) 2.0 software developed by \citet{boisse2012soap} and \citet{dumusque2014soap} to synthesize the observations into a time series of $n$ spectra of the full visible disk of the Sun. 
The power of SOAP 2.0 is that it is able to synthesize the spectra so as to accurately represent observations of an MS-G2V star under different stellar activity and viewing configurations, e.g., different spot sizes and {\edits locations}. This reconstruction of different situations is possible because the raw input spectra are spatially localized and precise, and the underlying geometry is relatively simple. 

To achieve our two goals of defining new stellar activity indicators and performing power-based model selection, we in fact use SOAP 2.0 multiple times. Firstly, we run SOAP 2.0 with a single large spot configuration to obtain a low-noise stellar activity dataset $Y_s$, an $n \times p$ matrix, {\edits where $n=125$ and  $p=237,944$}.  Each matrix row is a {\it p-}dimensional spectrum of the star at one of $n$ regularly spaced observation times.  In Section \ref{sec:pca_intro} below, we use $Y_s$ to {\it define} new high-information stellar activity indicators. 

Next, we use SOAP 2.0 to obtain many additional datasets for empirically computing exoplanet detection power, using the procedure described in Sections \ref{sec:selection_ideal} and \ref{sec:selection_practice}. In particular, we run SOAP 2.0 {\edits $M=1800$} more times with $M$ different stellar activity settings, plus adjustments to take account of the lower sign-to-noise (SNR) and irregular observation cadences typical in exoplanet surveys. The result is a collection, $Y_A$, of $M$  observation time vectors together with their corresponding data matrices, i.e.,  $Y_A =\left\{\left(t^{(m)},Y^{(m)}\right):m=1,\dots,M\right\}$, where each value of $m$ corresponds to  a different spot size and latitude setting. 
Further details {\edits regarding SOAP 2.0 and the datasets 
$Y_s$ and $Y_A$ are given in Section \ref{sec:data}.}

\subsection{Defining new stellar activity indicators}\label{sec:pca_intro} 
Since the Sun can be observed precisely, the dataset $Y_s$ is almost noiseless and we use it to {\it define} new stellar activity indicators. 
In spectroscopic surveys, a Doppler shift of the spectra of a host star is the only observable impact of an exoplanet. We therefore seek to separate this effect from the effects of stellar activity. Since stellar activity can also cause an apparent Doppler shift, it is not possible to immediately separate exoplanet signals from all stellar activity. However, stellar activity {\it can} be isolated by considering variations in the stellar spectra that are orthogonal to a Doppler shift. To summarize such variations, we apply a modified principal component analysis (PCA) to the data matrix $Y_s$, as detailed in Algorithm 1 below. 
\vspace{0.2cm}

\begin{figure}[t]
\includegraphics[width=1\textwidth,trim=5mm 20mm 5mm 0mm,clip]{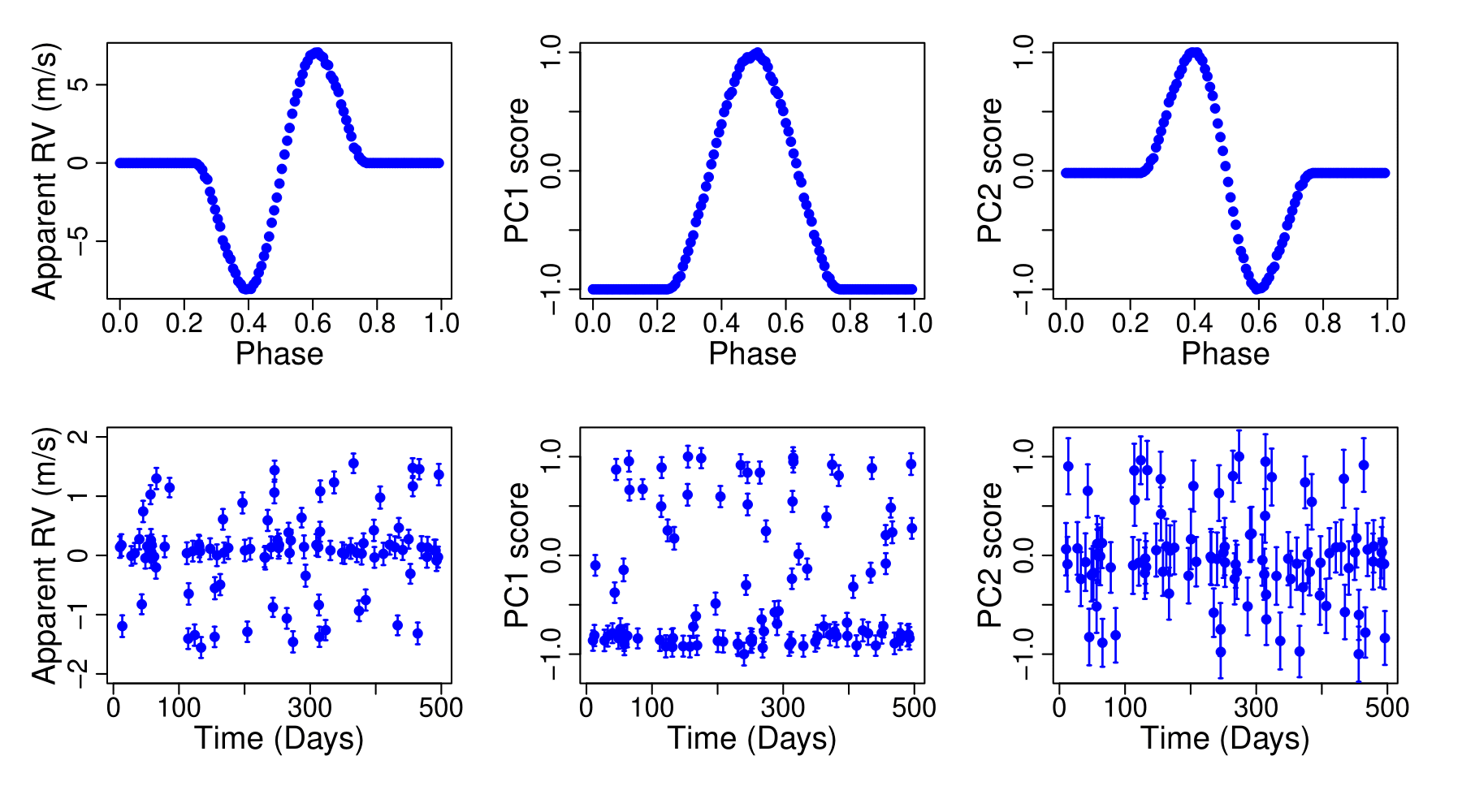} 
\caption{
Top row: noiseless realizations of $u$, $q_{\text{PC}1}$, and $q_{\text{PC}2}$ obtained by projecting $Y_s$ onto the components of $\mathcal{I}_{\text{\text{MS-G2V}}}$. 
Bottom row: realizations of $u$, $q_{\text{PC}1}$, $q_{\text{PC}2}$ for one of the noisy datasets contained in $Y_A$, plotted over time as opposed to in phase. 
\label{fig:gpca_outputs}}
\end{figure}

\noindent{\bf Algorithm 1}: Doppler-constrained PCA 
\begin{enumerate}
\item Input: $n\times p$ data matrix $Y_s$. 
    \item Based on $Y_s$, approximate the component $\boldsymbol{w}$ corresponding to a Doppler shift of a spectrum, see Appendix \ref{app:dopplerw}. 
    \item Subtract the Doppler component from the observed spectra:
    \begin{align}
\widetilde{Y}_s = Y_s - \frac{Y_s\boldsymbol{w}\boldsymbol{w}^T}{\sum_{i=1}^p |\boldsymbol{w}_i|^2},
\label{eqn:overview_remove_doppler}
\end{align}
    \item Apply PCA to $\widetilde{Y}_s$ to obtain the basis vectors $\boldsymbol{v}_k$, for $k=1,\dots,n-1$.
\end{enumerate}
The details of Algorithm 1 are discussed in Section \ref{sec:gpca}. The collection $\mathcal{I}_{\text{\tiny MS-G2V}}=\{\boldsymbol{w},\boldsymbol{v}_1,\dots,\boldsymbol{v}_\indnum\}$ defines $L$ stellar activity indicators plus the RV signal, where $L\leq n-1$ must be chosen. Given a dataset, the stellar activity indicators (and RV signal) are obtained by projecting onto the vectors in $\mathcal{I}_{\text{\tiny MS-G2V}}$, i.e., the indicators are the scores which we denote by $u(t_i)$ and $q_{\text{PC}j}(t_i)$, for $j=1,\dots,\indnum$  (here $u$ corresponds to $\boldsymbol{w}$). The indicators for $Y_s$ are plotted in the top row of Figure \ref{fig:gpca_outputs}. 
The scores $q_{\text{PC}j}(t_i)$, for $j=1,\dots,\indnum$, will always summarize only stellar activity. The scores $u(t_i)$ summarize either (a) stellar activity {\it and} a planet signal (if there is a planet); or (b) only stellar activity (if there is not a planet). In the case of Figure \ref{fig:gpca_outputs} there is in fact no planet.  For convenience, we write $\boldsymbol{s}=(\boldsymbol{u},\boldsymbol{q}_{\text{PC}1},\dots,\boldsymbol{q}_{\text{PC}\indnum})$, where $\boldsymbol{u}=(u(t_1),\dots,u(t_n))^T$ and
$\boldsymbol{q}_{\text{PC}j}=(q_{\text{PC}j}(t_1),\dots,q_{\text{PC}j}(t_n))^T$, for $j=1,\dots,\indnum$, and refer to $\boldsymbol{s}$ as the {\it stellar activity indicators}. 

{\edits The indicators defined by $\mathcal{I}_{\text{\tiny MS-G2V}}$ were derived using the specific spot size and latitude appearing in $Y_s$, but we found them to be almost identical for other configurations, suggesting that these indicators are generally useful, at least in the case of MS-G2V stars, which tend to have similar patterns of stellar activity.}  
{\edits In other words, for a new dataset from an MS-G2V,} it is reasonable to expect that its stellar activity will again be well captured by the stellar activity indicator components  $\mathcal{I}_{\text{\tiny MS-G2V}}$.
Conventional stellar activity indicators are similarly defined based on features of a low-noise or ideal spectrum, but they are typically not constructed in a way that necessarily extracts maximal information or even ensures that the indicators capture different information to each other. 

\subsection{Ideal power-based model selection}
\label{sec:selection_ideal}

Stellar activity is highly complex and it is generally challenging to jointly model multiple stellar activity indicators. We therefore propose a flexible class of models in Section \ref{sec:model_class}  and then use a model selection approach to find the best models within the class. For now, we focus on our model selection approach and postpone specification of the model class until Section \ref{sec:methodology}. 

{\edits Heuristically, in the context of  exoplanet detection, the point of stellar activity modeling is to  infer how much of the apparent RV signal shown in the top left panel of Figure \ref{fig:gpca_outputs} is due to stellar activity by making use of the two indicator time series plotted in the top middle and  top right panels.}
In other words, simply modeling the stellar activity data is not the primary objective, our goal is to {\it detect exoplanets via a hypothesis test}. The null hypothesis of the test is that there is no planet and a stellar activity model can sufficiently capture the data, and the alternative hypothesis is that there is a planet and therefore an additional model component is needed to capture the planet.  
Given our goal of planet detection, the correct way to select the model is to evaluate the planet detection power under each candidate stellar activity model, and then select the model yielding the highest power. This is what Algorithm 2 below does. For convenience we split the collection $Y_A$ into two parts $Y_N$ ($M_N=1000$ datasets) and $Y_P$ ($M_P=800$ datasets), which are used for generating a null distribution and generating datasets with planet signals, respectively.




\vspace{0.2cm}

\noindent{\bf Algorithm 2}: power-based model selection
\begin{enumerate}
\item Input: set $\mathcal{L}$ of candidate stellar activity models, SOAP 2.0 dataset collection $Y_A=(Y_N,Y_P)$, vector $(K_1,\dots,K_R)$ of planet signal magnitudes (in m/s), Type I error probability $\alpha$.
\item Compute null distribution and critical value
\begin{enumerate}[(i)]
\item Compute activity indicators $\boldsymbol{s}_N^{(m)}$ from $Y_N^{(m)}$, for $m=1,\dots,M_N$.
    \item For each $l\in \mathcal{L}$, compute the  likelihood ratio test (LRT) statistic
\begin{align}
 \Delta(l,\boldsymbol{s}_N^{(m)}) =  l_{\text{act}}(\hat{\theta}_{\text{act}}^{(m)};\boldsymbol{s}_N^{(m)}) -  l_{\text{full}}(\hat{\theta}_{\text{full}}^{(m)};\boldsymbol{s}_N^{(m)}),\label{eqn:lrt_stat}
\end{align}
for $m=1,\dots,M_N$, and let $\mathcal{F}_l$ denote the resulting distribution. Here $l_{\text{act}}$ and $l_{\text{full}}$ are the likelihood functions under the {\it stellar activity only model} and the {\it full model}, which additionally includes the planet component given by (\ref{eqn:planet}) in Section \ref{sec:planet_model}.
\item For each $l\in \mathcal{L}$, set $c_l$ to be the $1-\alpha$ quantile of $\mathcal{F}_l$. 
\end{enumerate}
\item Compute power
\begin{enumerate}[(i)]
\item Compute activity indicators $\boldsymbol{s}_P^{(m)}=(\boldsymbol{u}^{(m)},\boldsymbol{q}_{\text{PC}1}^{(m)},\dots,\boldsymbol{q}_{\text{PC}\indnum}^{(m)})$ from $Y_P^{(m)}$, for $m=1,\dots,M_P$.
    \item For $r \in \{1,\dots,R\}$, inject planet RV signal of magnitude $K_r$ m/s into $\boldsymbol{u}^{(m)}$, for $m=(r-1)n_p+1,\dots,rn_P$, where $n_P = M_P/R$.
    \item For each $l\in \mathcal{L}$, compute the empirical power for each value of $r$
    \begin{align}
       p(l,r) =  \frac{1}{n_P}\sum_{m=(r-1)n_p+1}^{rn_P}1_{\{\Delta(l,\boldsymbol{s}_P^{(m)}) > c_l\}}.\label{empirical_power}
    \end{align}.
\end{enumerate}
\vspace{-0.5cm}
\item Select $l\in \mathcal{L}$ that has the smallest value of $r$ satisfying $p(l,r)\geq 0.5$. 
\end{enumerate}
Note that, since each dataset has a different spot size and latitude, the power computed in (\ref{empirical_power}) marginalizes over these configuration settings. This is important because the configuration is unknown in practice.  

\subsection{Practical power-based model selection} 
\label{sec:selection_practice}
\vspace{0.2cm}

Unfortunately, Algorithm 2 cannot be applied directly in practice, because the number of candidate models is very large. Instead, in Algorithm 3 below, we first short-list some stellar activity models using three standard model selection criteria: the  Akaike information criterion (AIC), the Bayesian information criterion (BIC), and a  cross validation score (CV). The CV score we use is based on leaving out multiple observations, and the details are given in Appendix \ref{app:CV}. To construct the short-list we make use of $M_R=10$ datasets $Y_R^{(m)}$, $m=1,\dots,M_R$, that well represent the distribution of spot sizes and latitudes. These can either be selected from $Y_A$ or simulated from SOAP 2.0 in the same manner. Here we use the latter approach and choose the representative latitudes and sizes using a maximin Latin hypercube design.  
\vspace{0.2cm}

\noindent{\bf Algorithm 3}: power-based model selection in practice
\begin{enumerate}
    \item Input: set $\mathcal{L}$ of candidate stellar activity models, datasets $=Y_R^{(m)}$, for $m=1,\dots,M_R$. 
    \item Compute activity indicators $\boldsymbol{s}_R^{(m)}$ from $Y_R^{(m)}$, for $m=1,\dots,M_R$.
    \item For each $l\in \mathcal{L}$, compute 
    \begin{align}
        \overline{\Delta\text{AIC}}(l) = \frac{1}{M_R}\sum_{m=1}^{M_R} \text{AIC}(l,\boldsymbol{s}_R^{(m)}) - \underset{k\in \mathcal{L}}{\text{min}}\text{AIC}(k,\boldsymbol{s}_R^{(m)}),\label{eqn:average_crit}
    \end{align}
    and similarly compute $\overline{\Delta\text{BIC}}(l)$ and $\overline{\Delta\text{CV}}(l)$.
    \item Short-list the top 5 models under each of the 3 selection criteria.
    \item Apply Algorithm 2 to the short-list of models. 
\end{enumerate}
In (\ref{eqn:average_crit}),  $\text{AIC}(l,\boldsymbol{s}_R^{(m)})$ denotes the AIC for model $l$ and dataest $\boldsymbol{s}_R^{(m)}$. The results of applying Steps 1-4 of Algorithm 3 are presented in Table \ref{tab:activity_model_results} in Section \ref{sec:preliminary_selection}.
The final model selected is denoted $\mathcal{M}_{\text{\tiny MS-G2V}}$.

For a new dataset from an MS-G2V star, it is reasonable to expect that its stellar activity will again be well captured by the  indicators defined by $\mathcal{I}_{\text{\tiny MS-G2V}}$  and the corresponding 
model $\mathcal{M}_{\text{\tiny MS-G2V}}$ identified by Algorithm 3. 
Thus, in principle, no new indicator and model development is needed when applying our method to a new dataset (from an MS-G2V star): we simply compute the stellar activity indicators (using $\mathcal{I}_{\text{\tiny MS-G2V}}$) and perform a likelihood ratio test. The null distribution used by the test is again that previously found in Step 3 of Algorithm 2. 

{\edits Nonetheless, given the possibility of more complex variations in the form of activity across MS-G2V stars, a case for re-computing the indicator components $\mathcal{I}_{\text{\tiny MS-G2V}}$ for each new dataset could be made, and this approach is illustrated in Section \ref{sec:generalizing}. However, except in the presence of very large spots, we found that principal components computed from noisy datasets were generally noisy themselves, and less useful for exoplanet detection than  $\mathcal{I}_{\text{\tiny MS-G2V}}$. Furthermore, re-computing the PCA basis for each dataset adds substantial computation to Algorithm 3. Therefore, in the main part of this paper, we  use the fixed indicators given by $\mathcal{I}_{\text{\tiny MS-G2V}}$ and the model $\mathcal{M}_{\text{\tiny MS-G2V}}$. 
In this respect our method is similar to conventional approaches 
which pre-define indicators and models  based on physical knowledge. The difference is that we derive the indicators empirically (from low-noise data), and select a model from a large class based on power for exoplanet detection.} 

\section{Spectroscopic Time Series Datasets}
\label{sec:data}

\subsection{Dataset $Y_s$ used to define stellar activity indicators}
\label{sec:soap}

The low-noise dataset $Y_s$ is an $n\times p$ matrix containing a times series of $n=125$ spectra each of dimension $p=237,944$, i.e., each of the $n$ spectra consist of the measured light intensity at the same $237,944$ wavelengths. 
The dataset captures stellar activity in the form of a single stellar spot and is used as an ideal dataset with respect to which we  define new stellar activity indicators by running Algorithm 1 introduced in Section \ref{sec:pca_intro}, see Section \ref{sec:pca_method} for full details.
To obtain $Y_s$ we ran SOAP 2.0 a single time to synthesize the {\edits raw input spectra described in Section \ref{sec:data_overview} 
} into a time series of spectra of the full visible disk of the Sun.  The observation times are equally spaced and have Solar rotation phases $p_i = i/125$, for $i=0,\dots,124$. For our SOAP 2.0 settings, the activity takes the form of a single {\edits large} spot that covers 1\% of one hemisphere of the Solar surface, the spot latitude from the Solar equator is $40\degree$, and the inclination of the Solar rotation axis relative to the line of sight is $90\degree$. 

Unfortunately, {\edits it is not practical to observe stars more distant than the Sun with the same resolution and signal-to-noise} as our raw input data. Thus, we modify SOAP 2.0 to reduce the output resolution to a practically achievable level. The result is that the number of wavelengths recorded is reduced from $523,732$ (in the raw input spectra) to $p=237,944$.\footnote{More precisely, the resolution is reduced from $\sim10^4$ to $1.5 \times 10^5$ 
by convolving the higher resolution spectra with a Gaussian line spread function and resampling using cubic splines.} This modification is necessary because if we were to define stellar activity indicators based on {\edits unrealistically} high resolution spectra, then the indicators may primarily capture spectral features that would not be observable in practice. 




\subsection{Dataset $Y_A$ used to perform detection power-based model selection}\label{sec:adjustments}

The dataset $Y_A$ is a collection of observations from $M=1800$ MS-G2V stars generated by SOAP 2.0. For each star, $Y_A$ contains a vector of  $n_A=100$ observation times and a corresponding data matrix containing $n_A$ spectra, i.e.,  $Y_A =\left\{\left(t^{(m)},Y_{n_A \times p}^{(m)}\right):m=1,\dots,M\right\}$. 
The dataset is representative of exoplanet survey observations of {\edits $1800$} MS-G2V stars with varying degrees of stellar activity and Algorithm 3 uses it to perform model selection by empirically evaluating  planet detection power under different stellar activity models, see Section \ref{sec:model_selection_procedure}.  To obtain $Y_A$ we ran SOAP 2.0 $M$ times, where 
for each run the spot size and latitude were drawn from the distribution of configurations expected for the Sun (or similar stars). Specifically, the distributions are
\begin{align}
    S_{\text{lat}} &\sim \pi\mathcal{N}_{ab}(\mu,\sigma^2)+(1-\pi)\mathcal{N}_{ab}(-\mu,\sigma^2) \label{eqn:lat}\\
        S_{\text{size}} &\sim \rho\mathcal{LN}_{d}(\theta_1,\gamma^2)+(1-\rho)\mathcal{LN}_{d}(\theta_2,\gamma_2^2) \label{eqn:size},
\end{align}
where $\mathcal{N}_{ab}$ denotes a truncated Gaussian distribution with $a=-90\degree$ and $b=90\degree$ being the lower lower and upper bound, respectively, and  $\mathcal{LN}_{d}$ denotes a lower truncated log-Gaussian distribution with lower bound $d=10$ micro Solar hemispheres (MSH). To make  (\ref{eqn:lat}) and (\ref{eqn:size}) consistent with the distributions given in  \citet{baumann} and \citet{mandal}, respectively, we set $\pi=0.5$, $\mu=15.1$, $\sigma=7.3$, and $\rho=0.4$, $\theta_1=\log(46.51)+\log(2.14)$, $\theta_2=\log(90.24)+\log(2.49)$, $\gamma_1=\sqrt{\log(2.14)}$, $\gamma_2=\sqrt{\log(2.49)}$ (where $\log$ denotes the natural logarithm). Although in practice there may be multiple spots, they typically form in localized groups which can be approximated as a single spot (as is done here), and the distribution \eqref{eqn:size} is consistent with this approximation. 

To ensure that the detection powers computed by Algorithm 3 are relevant, we set the observation cadence and  signal-to-noise (SNR) to be reasonable for exoplanet surveys.
Specifically,  for each star (i.e., each value of $m$), we set the total observation window to be 500 days. 
Next,  we randomly select $n_{A}=100$ phases $p^{(m)}_1,\dots,p^{(m)}_{100},$ from among $p_i=i/125$, $i=0,\dots,124$ (without replacement).
Finally, 
we randomly generate a rotation cycle of the star $r^{(m)}_i \in \{0,\dots,49\}$ and set  the observation time to be $t^{(m)}_i=(r^{(m)}_i+p^{(m)}_i)\tau^{(m)}_{s}$, for $i=1,\dots,n_A$, where $\tau_s$ is the stellar rotation period. In this work we set $\tau^{(m)}_s=\tau_s=10$ days.
To generate spectra with a given SNR, we add noise  $\epsilon_{t\lambda}$ to the spectrum at each wavelength.
Specifically,
\begin{align}
    \epsilon_{t\lambda} \sim \mathcal{N}(0,\beta_t f_t(\lambda)/\text{SNR}) \label{eqn:noise}
\end{align}
where $\beta_t = \frac{1}{\lambda_{\text{max}}-\lambda_{\text{min}}} \int^{\lambda_{\text{max}}}_{\lambda_{\text{min}}} f_t(\lambda) d\lambda$.
This noise distribution is based on the fact that a Poisson distribution accurately models the uncertainty involved in the physical process of photons being detected by a telescope. A Gaussian is used in (\ref{eqn:noise}) because $f_t(\lambda)$ is continuous due to observation stage processing steps. 
We set SNR=500, unless otherwise specified. This is higher than typical of existing observations, but still feasible  for exoplanet surveys.

\section{Defining New Stellar Activity Indicators}
\label{sec:gpca}

\subsection{Spectroscopic stellar activity indicators}
\label{sec:indicators}

It is prohibitively challenging to directly model the temporal evolution of the high-dimensional spectra stored in the rows of the data {\edits matrix $Y_s$ (or $Y_A^{(m)}$, for $m=1,\dots,M$)}. Instead astronomers typically model time series of a small number of stellar activity indicators, which are functionals of individual spectra summarizing the level and nature of stellar activity. 
The indicators used by \citet{rajpaul2015gaussian} (hereafter \citetalias{rajpaul2015gaussian}) are $\logrhk$ and BIS, which measure the emission in the core of the Calcium II H \& K spectral lines and the asymmetry of an ``averaged'' version of the spectral lines, respectively. 
It is non-trivial to compute $\logrhk$, and \citetalias{rajpaul2015gaussian} suggest replacing it by normalized flux when $\logrhk$ is not available.  

Some stellar activity indicators are summaries of very specific parts of the stellar spectrum, e.g., $\logrhk$. More generally, a stellar activity indicator can be any functional $g$ of the stellar spectra observed. 
This raises the question of how to choose $g$ in order to capture as much information as possible. Indeed, the measures $\logrhk$ and BIS were designed for purposes other than planet detection, and therefore in the current context are somewhat arbitrary. 
It is thus desirable to have {\edits an empirical} method for identifying stellar activity indicators that are informative for our goals of detecting and characterizing exoplanets.

\subsection{Defining new stellar activity indicators via Doppler-constrained PCA}
\label{sec:pca_method}

Recently, \citet{davis2017insights} proposed using principal component analysis (PCA) to construct stellar activity indicators in {\edits an empirical} way. However, it is not immediately apparent how their stellar activity indicators can be used in practice because it is unclear if any given principal component corresponds to stellar activity, a planet RV signal, or both. 

Algorithm 1 in Section \ref{sec:pca_intro} specifies our simple adaptation of PCA which separates the apparent RV signal from the PCA scores  and thereby addresses the difficulties posed by the \citet{davis2017insights} indicators. In summary, PCA is applied to $\widetilde{Y}_s = Y_s - {Y_s\boldsymbol{w}\boldsymbol{w}^T}/{\sum_{i=1}^p |\boldsymbol{w}_i|^2}$, where  the $p \times 1$ component $\boldsymbol{w}$  corresponds to a Doppler shift, and is specified below. 
We denote the PCA components by $\boldsymbol{v}_j$, for $j=1,\dots,n-1$. 
The collection of basis vectors $\mathcal{I}_{\text{\tiny MS-G2V}}=\{\boldsymbol{w},\boldsymbol{v}_1,\dots,\boldsymbol{v}_l\}$ define the apparent RV signal and our stellar activity indicators for MS-G2V stars. Based on the findings of \citet{davis2017insights} and our own experience, $L$ does not need to be larger than 2 or 3, so we choose $\indnum=2$. 

Since  Doppler shifts of the spectral wavelengths are very small, they 
closely correspond to shifts in the direction of the spectrum derivative. We therefore choose $\boldsymbol{w}$ in Algorithm 1 to be the derivative of the average spectrum in $Y_s$, i.e., $\boldsymbol{w}=(\bar{f}'(\lambda_1),\dots,\bar{f}'(\lambda_p))^T$, see Appendix \ref{app:dopplerw} for full details.
We emphasize that $\boldsymbol{w}$ is simply the direction along which a Doppler shift occurs, not the magnitude or sign of an observed Doppler shift. Therefore, even if there was a planet in the dataset used to define $\boldsymbol{w}$ (which for $Y_s$ there is not) then this would not 
change $\boldsymbol{w}$  significantly. 

The bottom row of Figure \ref{fig:gpca_outputs} shows our stellar activity indicators for one of the stars in the cadence and SNR adjusted dataset $Y_A$. Specifically, we have plotted $u(t_i)$ and $q_{\text{PC}j}(t_i)$, for $j=1,2$ and $i=1,\dots,100$. Two standard deviation error bars are plotted to indicate the uncertainty induced by the spectrum level noise, as described by (\ref{eqn:noise}).

\section{Framework for modeling planet signals and stellar activity}
\label{sec:methodology}

\subsection{Keplerian model for planetary RV signals}
\label{sec:planet_model}

The RV signal due to a single planet orbiting a star is well understood and can be described  precisely using a Keplerian  model, see for example \citet{danby1988fundamentals}. Specifically, the RV induced by a single planet system is given by
\begin{align}
v(t) & = K(e\cos \omega + \cos(\omega + \phi(t))) + \gamma,
\label{eqn:planet}
\end{align}
where $e$ is the eccentricity of the planetary orbit, $\omega$ is an orbital orientation angle  known as the {\it argument of periapsis}, and $\gamma$ and $K$ are velocity offset and amplitude parameters, respectively. 
The angle $\phi(t)$ is called the {\it true anomaly} and indicates the phase of the star in its elliptical orbit of the center of mass. This angle is determined by a system of three equations which are given in Appendix \ref{app:planet} and which depend on the planet orbital period $\tau_p$, {\edits in addition to other parameters.} 
The right panel of Figure \ref{fig:spectrum} shows an example planet RV signal (solid line). 


\subsection{General class of stellar activity models}
\label{sec:model_class}

We now propose a class of Gaussian process models to jointly capture changes in the RV corruption due to stellar activity and times series of multiple stellar activity indicators, {\edits e.g., those in the bottom row of Figure \ref{fig:gpca_outputs}}. The indicators can be general and need not be the Doppler-constrained PCA scores discussed in Section \ref{sec:pca_method}. 
Our approach is an extension of the model proposed by \citetalias{rajpaul2015gaussian}. 
{\edits More generally, our model class can be viewed as an adaptation of the linear model of co-regionalization (LMC), see for example \citet{journel1978mining}, \citet{osborne2008towards}, and \citet{alvarez2011computationally}.}

A real-valued continuous time stochastic process $\{\gp(t)\}$ is a Gaussian process (GP) if for every finite set of times $t_1,\dots,t_n$ the vector $(\gp(t_1),\dots,\gp(t_n))$ has a multivariate Gaussian distribution. 
Here we make the usual assumptions that for any $t$  the mean of $X(t)$ is zero and the process is stationary in time so that the covariance between two observations of the process $\gp(t)$ and $\gp(t')$ only depends on the value of $|t-t'|$. Specifically, we adopt the quasi-periodic covariance function 
\begin{align}
\label{eqn:kernel}
\kernel(t,t') = \exp\left(-\frac{\sin^2(\pi (t-t')/\tau_s)}{2 \lambda_p^2} - \frac{(t-t')^2}{2\lambda_e^2}\right),
\end{align}
where $\tau_s$ is the stellar period and $\lambda_p$ and $\lambda_e$ are parameters governing the relative importance of periodic and local correlations and the time-scale of local correlations, respectively. That this covariance function leads to a positive definite covariance matrix follows from the fact that the product of two valid covariance functions yields a valid covariance function, see \citet{rasmussen2006gaussian} page 95. 
The same covariance function was also adopted by \citetalias{rajpaul2015gaussian}. 
The motivation for a quasi-periodic covariance structure is that for a given spot the stellar activity signal should be similar for each rotation of the star but will change over longer intervals due to evolution of the spot or other phenomena not explicitly modeled. We denote the parameters of the covariance function by $\phi=(\tau_s,\lambda_p,\lambda_e)$. The reader is referred to \citet{rasmussen2006gaussian} for a comprehensive introduction to Gaussian processes. 

We also make use of the first two derivatives of $\gp(t)$ in our model class (\ref{eqn:model1})-(\ref{eqn:our_errors}) below, where we have denoted them by $\dot{\gp}(t)$ and $\ddot{\gp}(t)$. The inclusion of $\dot{X}(t)$ is motivated by physical arguments in \citet{aigrain2012simple}, which demonstrate that  the RV corruption due to a spot can be approximated by a function of the area of the spot projected onto our line of sight and the derivative of this area. 
{\edits We additionally include $\ddot{X}(t)$ to capture second order effects of the spot area, as suggested by \citetalias{rajpaul2015gaussian} (though, for simplicity, they did not use this second derivative).} The derivative of a Gaussian process with covariance function $\kernel(t,t')$  (if it exists as is the case for (\ref{eqn:kernel})) is also a Gaussian process and has covariance function 
\begin{align*}
\frac{\partial^2}{\partial t \partial t'}\kernel(t,t').
\end{align*}
Furthermore, $\partial \kernel(t,t')/ \partial t'$ gives the covariance between $\gp(t)$ and $\dot{\gp}(t')$. The covariance function for higher order derivatives of $\gp(t)$ can be obtained in an analogous way. These results follow from Theorem 2.2.2 in \citet{adler2010geometry}. 

Let $u(t)$ and $q_j(t)$, for $j=1,\dots,\indnum$, denote the values of the RV corruption and $l$ generic stellar activity indicators, respectively, at time $t$. Given observation times $t_1,\dots,t_n$, we propose the following class of models:
\begin{align}
u(t_i) &= m_0 + a_{01}X(t_i) + a_{02}\dot{\gp}(t_i) + a_{03}\ddot{\gp}(t_i) + a_{04}\extragp_0(t_i)+ \epsilon_{0i}\label{eqn:model1}\\
q_1(t_i) &= m_1 + a_{11}X(t_i) + a_{12}\dot{\gp}(t_i) + a_{13}\ddot{\gp}(t_i) + a_{14}\extragp_1(t_i)+ \epsilon_{1i}\label{eqn:model2}\\
\vdots \nonumber\\
q_{\indnum}(t_i)  &= m_\indnum + a_{\indnum1}X(t_i) + a_{\indnum2}\dot{\gp}(t_i) + a_{\indnum3}\ddot{\gp}(t_i) + a_{\indnum4}\extragp_\indnum(t_i)+ \epsilon_{\indnum i},\label{eqn:model_class}
\end{align}
for $i=1,\dots,n$, where the $\epsilon_{ji} $ are independent with
\begin{align}
\epsilon_{ji} \sim \mathcal{N}(0,\sigma_{ji}^2),\label{eqn:our_errors}
\end{align}
for $j=0,\dots,\indnum$. Here the $\sigma_{ji}$ are given measurement uncertainties, the $a_{jk}$ are unknown parameters to be inferred from the data, and $Z_0,\dots,Z_\indnum$ are independent zero mean GPs with the covariance function (\ref{eqn:kernel}). The parameters for the covariance functions of $Z_0,\dots,Z_\indnum$ are  assumed to be the same and are denoted $\phi_Z$ (they are allowed to differ from $\phi$, the covariance parameters for $X$). 
We denote by $\Sigma$ the $(L+1)n\times (L+1)n$ covariance matrix implied by the model (\ref{eqn:model1})-(\ref{eqn:our_errors}), and specify its form in Appendix \ref{app:C}.

The purpose of the independent GP components $Z_0,\dots,Z_\indnum$ is to permit structured deviations from linear combinations of the physically motivated terms $X(t)$, $\dot{X}(t)$, and $\ddot{X}(t)$. \citetalias{rajpaul2015gaussian} suggested that {\edits their approach could perhaps be improved by adding in} these independent GP components, and indeed when we tried modeling a variety of stellar activity indicators using (\ref{eqn:model1})-(\ref{eqn:our_errors}) we found that this additional flexibility is sometimes helpful (despite our inclusion of $\ddot{X}(t)$). 


We now specify the log-likelihood function of the model (\ref{eqn:model1})-(\ref{eqn:our_errors}). 
For conciseness we write the observation times as $\boldsymbol{t}=(t_1,\dots,t_n)^T$ and the time series data as $\boldsymbol{s}=(\boldsymbol{u},\boldsymbol{q_1},\dots,\boldsymbol{q_\indnum})$, where $\boldsymbol{u}=(u(t_1),\dots,u(t_n))^T$ and
$\boldsymbol{q_j}=(q_j(t_1),\dots,q_j(t_n))^T$, for $j=1,\dots,\indnum$.  Denoting the parameters for the model (\ref{eqn:model1})-(\ref{eqn:model_class}) by $\theta_{\text{act}} =(m_0,\dots,m_\indnum,a_{01},\dots,a_{04},$  $\dots,a_{\indnum1},\dots,a_{\indnum4},\phi,\phi_Z)$, the log-likelihood is 
\begin{align}
l_{\text{act}}(\theta_{\text{act}}|\boldsymbol{t},\boldsymbol{s}) = -\frac{(L+1)n}{2}\log(2\pi) -\frac{1}{2}\left|\Sigma\right|-\frac{1}{2}(\boldsymbol{s}-\boldsymbol{m})^T\Sigma^{-1}(\boldsymbol{s}-\boldsymbol{m}),
\label{eqn:likelihood}
\end{align}
where 
$\boldsymbol{m}=(m_0\boldsymbol{1}_n^T,m_1\boldsymbol{1}_n^T,\dots,m_\indnum\boldsymbol{1}_n^T)$ and $\boldsymbol{1}_n$ denotes a vector of $n$ ones. 

Our model capturing both a planet and stellar activity is simply (\ref{eqn:model1})-(\ref{eqn:model_class}) except that  $u(t)$ in (\ref{eqn:model1}) is replaced by
\begin{align}
u_p(t) = u(t) + v(t) - \gamma,
\label{eqn:add_planet}
\end{align}
where  $v(t)$ is given by (\ref{eqn:planet})  and the offset $\gamma$  is subtracted out because $m_0$ (in (\ref{eqn:model1}))  already provides an offset. We refer to this model incorporating a planet as the {\it full model} and denote the corresponding log-likelihood by $l_{\text{full}}$. We write the parameters of the full model as $\theta_{\text{full}}= (\theta_{\text{act}},\alpha)$, where $\alpha=(K,M_0,\tau_p,\omega,e)$ are the parameters describing the planet and its orbit. 


\subsection{\citet{rajpaul2015gaussian} model as a special case}
\label{sec:rajpaul}

The model proposed by \citetalias{rajpaul2015gaussian}  for the RV corruption and the stellar activity indicators $\logrhk$ and $\bis$ is a special case contained within our model class (\ref{eqn:model1})-(\ref{eqn:our_errors}). Setting $q_1$ and $q_2$ to be the indicators $\logrhk$ and $\bis$, respectively, their model is
\begin{align}
u(t_i) &= m_0 + a_{01}\gp(t_i) + a_{02}\dot{\gp}(t_i) + \epsilon_{0i}\label{eqn:rajpaul1}\\
q_1(t_i) &= m_1 + a_{11}\gp(t_i)+ \epsilon_{1i}\label{eqn:rajpaul2}\\
q_2(t_i) &= m_2 + a_{21}\gp(t_i) + a_{22}\dot{\gp}(t_i)+ \epsilon_{2i}\label{eqn:rajpaul3}.
\end{align}
A minor difference from the model they specify is that we assume $\epsilon_{ji} \sim N(0,\sigma_{ij}^2)$, for $j=0,1,2$ and $i=1,\dots,n$, where the $\sigma_{ij}$ are the individual observation uncertainties recorded. In contrast, \citetalias{rajpaul2015gaussian} set $\sigma_{ij}=\sigma_j$,  and estimate $\sigma_j$, for $j=0,1,2$. We use the recorded uncertainties $\sigma_{ij}$ for all model fits in this paper (providing measurement uncertainties is standard practice in astronomy). 

We consider the approach by \citet{rajpaul2015gaussian} to be representative of the current state-of-the-art of stellar activity modeling because it is one of a few existing models in the literature that meets the key criteria for good stellar activity modeling performance  identified in a recent comparison of methods by \citet{dumusque2017radial}. 
Furthermore, the \citetalias{rajpaul2015gaussian} model (and slight variants) has proved useful for both detecting and characterizing low-mass exoplanets and for recognizing spurious claims of planets that were in fact due to stellar activity, see \citetalias{rajpaul2015gaussian} and \citet{rajpaul2015ghost}.  Therefore, we will use the \citetalias{rajpaul2015gaussian} model as a point of reference in evaluating our results in Section \ref{sec:selection}.

\section{power-based model selection}
\label{sec:model_selection_procedure}

If we set the stellar activity model to be (\ref{eqn:model1})-(\ref{eqn:our_errors}) allowing all the coefficients to be non-zero then the power of the resulting likelihood ratio test (LRT) for detecting a planet is {\edits sub-optimal} for realistic planets, because the stellar activity model is too flexible and absorbs planet signals. That is, the test statistic $\Delta=l_{\text{act}}(\hat{\theta}_{\text{act}}) -  l_{\text{full}}(\hat{\theta}_{\text{full}})$ 
tends to be small even when a planet is present. 
Thus, we cannot allow all the coefficients in (\ref{eqn:model1})-(\ref{eqn:model_class}) to be non-zero but must try to find the best sub-model for the specific stellar activity indicators at hand. 
 
The ideal way to choose between the models in the class (\ref{eqn:model1})-(\ref{eqn:our_errors}) is to apply Algorithm 2  in Section \ref{sec:selection_ideal} which selects the best model by comparing the power for planet detection under each candidate model. Unfortunately, since each coefficient $a_{jk}$ in  (\ref{eqn:model1})-(\ref{eqn:model_class}) can be included or not, there are $2^{12}=4096$ models, and the computational cost of evaluating the power for all of them is prohibitively high. Some of the models can be ruled out straight away, but there are still too many to apply Algorithm 2 directly. Instead, we apply Algorithm 3  in Section \ref{sec:selection_practice} which first creates a short-list of models using  the Akaike information criterion (AIC), the Bayesian information criterion (BIC), and cross validation (CV), and then applies Algorithm 2 to this short-list.

We now give additional details regarding Algorithm 3. In principle our procedure can be applied with any number of activity indicators, but for concreteness we fix $\indnum=2$. We can immediately exclude models in which $a_{j1}=a_{j2}=a_{j3}=a_{j4}=0$ for any $j\in\{0,1,2\}$ because a non-trivial model is required for each time series. We also fix $a_{04}=0$ because including an independent GP for modeling the RV corruption  will intuitively lead to {\edits low power; the model would be able to capture RV signals from planets without the need for a planet model.}  
This leaves $(2^3-1)(2^4-1)^2=1575$ possible stellar activity models from which we create a short-list based on the ability of the models to capture the stellar activity indicators $u(t_i)$ and $q_{\text{PC}j}(t_i)$, for $j=1,2$ and $i=1,\dots,n$, plotted in the bottom row of Figure \ref{fig:gpca_outputs}. 

In Steps 3 and 4 of Algorithm 3, we {\edits screen} the models and create the short-list by {\edits computing the   criteria $\overline{\Delta\text{AIC}}$, $\overline{\Delta\text{BIC}}$, and $\overline{\Delta\text{CV}}$, which average across 10 screening datasets}, and are specified by \eqref{eqn:average_crit}. The specifics of our CV procedure are given in Appendix \ref{app:CV}. 
For the sake of illustration we include the best five models under each criterion in the short-list. 


Step 5 of Algorithm 3 applies Algorithm 2 to assess the planet detection power under the short-listed models. 
For  each model, Step 2 of Algorithm 2 approximates the null distribution of the LRT statistic  $\Delta$ by computing it for the 1000 replicate datasets stored in $Y_N$ (recall that $Y_A=(Y_N,Y_P)$). Details of the optimization approach used are given in Appendix \ref{app:D}. Step 3(ii) of Algorithm 2, injects  planet signals into the {\edits 800} replicate datasets stored in $Y_P$. In practice, the specific null and planet-injected datasets that should be used in the power approximations will depend on the type of activity expected, the instrument in question, our understanding of what signals can realistically be detected, and the type of planets we are hunting.  In this paper, we 
inject planets with an orbital period of 7 days and RV signal amplitudes close to or below the current detection threshold of $\sim$ 1 m/s. The 7 day period is favorable in the sense that it does not coincide with the stellar rotational period in our dataset (10 days) or its harmonics. To inject a planet signal we simply add $v(t_i;\alpha)$ (given by  (\ref{eqn:planet})) to $u(t_i)$, for $i=1,\dots,n$, and remove the superfluous offset term, see (\ref{eqn:add_planet}). The parameter vectors of the injected signals are given by $\alpha = (K,M_0=1.5,\tau_p=7,\omega=1,e=0.2,\gamma=0)$ for amplitudes {\edits $K=0.01,0.02,\dots,0.16$ m/s}. Planet signals corresponding to each setting of $\alpha$ are injected into 50 replicate datasets. In practice, the parameters $M_0$, $\omega$, $e$, and $\gamma$ have less impact on the planet signal than $K$ and $\tau_p$ and the fixed values used here are reasonably typical. Note that {\it all} the planet parameters are fitted in our procedure. 

In Step 3(iii) of Algorithm 2, for each model $l\in \mathcal{L}$, we compute $\Delta$ for each of the {\edits $800$} planet-injected datasets. For each dataset, we reject the null of no planet if $\Delta$ is greater than the 0.99 quantile (denoted $c_l$ in Algorithm 2) of the relevant null distribution, i.e., that constructed using the stellar activity model $l$. The number of rejections across the 50 simulations at a given value of $K$ provides an estimate $p(l,K)$ of the detection power at that planet signal amplitude under model $l$, see \eqref{empirical_power}. 

Step 4 of Algorithm 2 selects the final model $\mathcal{M}_{\text{MS-G2V}}$ to be that with the smallest {\it detection threshold}, which we define to be the minimum amplitude of a planet RV signal $K$ such that $p(l,K)\geq 0.5$. 
If no activity model dominated the others in terms of detection power for the simulated planet configurations, we would choose the final model according to additional considerations, e.g., {\edits the best overall criterion rank (i.e., the minimum value of the sum of the ranks under the three selection criteria)}, the highest detection power for the planets of most interest, or the planets that are deemed most probable to accompany the type of star in question.

\section{Application of model selection procedure}
\label{sec:selection}

\subsection{Preliminary model}
\label{sec:preliminary}

\begin{figure}[t]
\centering
\includegraphics[width=0.85\linewidth,trim=0mm 0mm 0mm 10mm,clip]{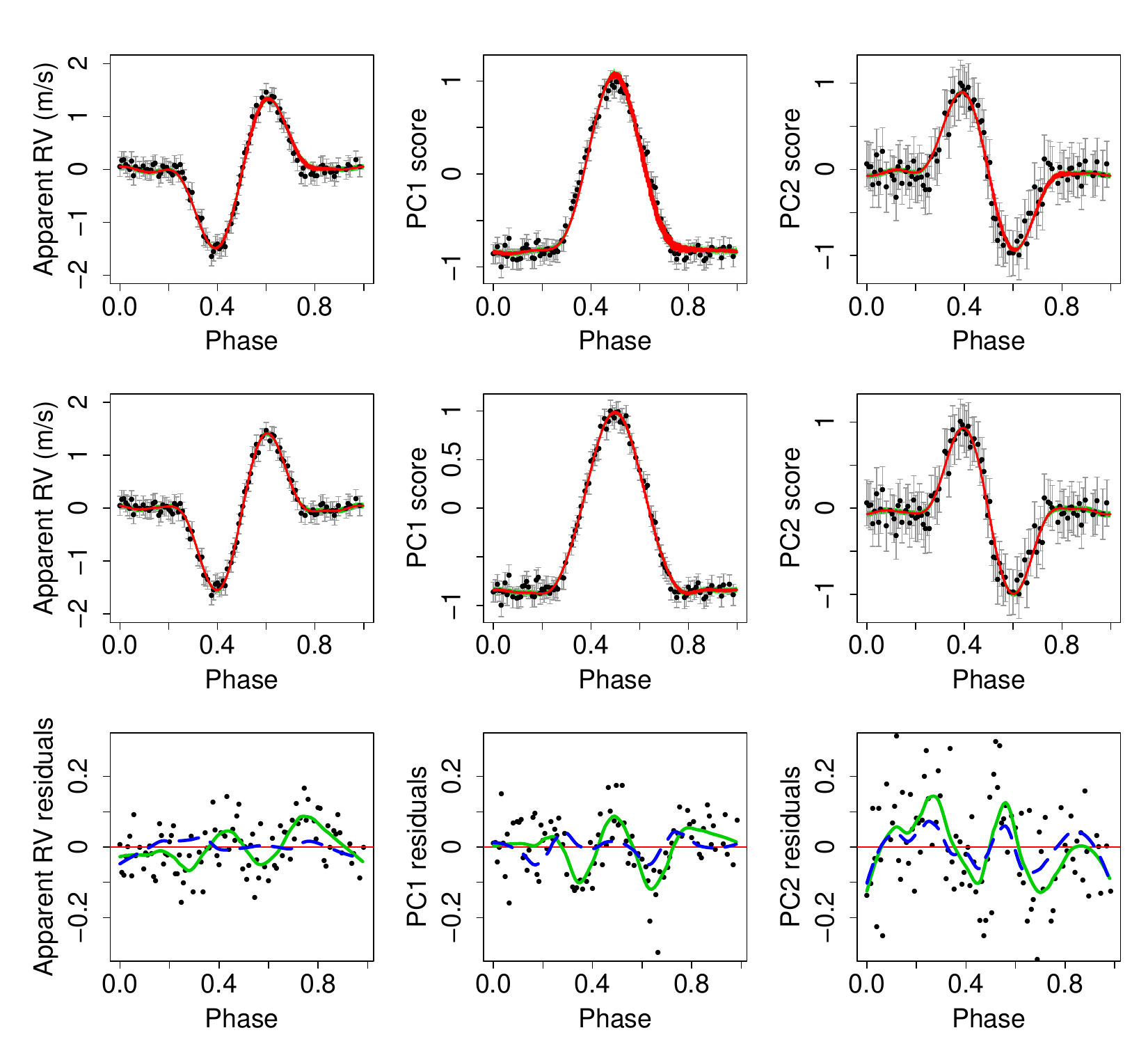}
\caption{
Preliminary model fit to $\{u(t_i),q_{\text{PC}1}(t_i),q_{\text{PC}2}(t_i)\}_{i=1,\dots,n}$ (top row) and AIC-1   model fit (middle row). Note that during fitting all the signals were normalized for numerical stability but $u(t)$ is plotted on the original m/s scale for interpretability. The bottom row shows the  residuals for the preliminary model fit (points), the solid green curve is a LOESS (local polynomial regression) fit to the plotted residuals, and the dashed blue line is a LOESS fit to the AIC-1 model residuals (which are not plotted). \label{fig:model_fits}}
\end{figure}

The \citetalias{rajpaul2015gaussian} model  (\ref{eqn:rajpaul1})-(\ref{eqn:rajpaul3}) was not designed to capture our PCA based stellar activity indicators because our indicators differ to $\logrhk$ and BIS. However, the plot of $\{u(t_i),\logrhk(t_i),\text{BIS}(t_i)\}_{i=1,\dots,m}$ in Figure 3 of \citetalias{rajpaul2015gaussian} qualitatively resembles the top row of Figure \ref{fig:gpca_outputs} here which shows $\{u(t_i),q_{\text{PC}1}(t_i),q_{\text{PC}2}(t_i)\}_{i=1,\dots,n}$ for the noiseless dataset $Y_s$. To investigate further, in the top row of Figure \ref{fig:model_fits} we plot the maximum likelihood fit of the \citetalias{rajpaul2015gaussian} model to {\edits the noisy dataset shown in the 
bottom} row of Figure \ref{fig:gpca_outputs}. {\edits For visualization purposes, the $x$-axis is stellar rotation phase rather than time, though the stellar rotation period was treated as unknown when fitting the model.} 
A lighter region representing a 95\% confidence interval for the underlying functions is plotted but is mostly covered by the fitted function (bold line) due to very small uncertainties. The reason for the small uncertainties is that  $X(t)$ (and hence $\dot{X}(t)$) can be precisely inferred by combining information across the three time series. 
We will refer to the \citetalias{rajpaul2015gaussian} model as the {\it preliminary model} since it is our best initial guess of a suitable model for our indicators based on the existing literature. 

Despite the high precision, the preliminary model fit is unsatisfactory because  the inferred underlying function shows systematic deviations from the observations. This can be seen from the bottom row of  Figure \ref{fig:model_fits}, which shows the residuals for the preliminary model fit, {\edits with the solid green lines being  LOESS (local linear regression) fits to the plotted residuals.} For instance, consider the middle panel of the bottom row which gives the $q_{\text{PC}1}$ residuals: we can see {\edits oscillations, with the residuals being positive near phase $0.5$, and negative either side of $0.5$.} {\edits Furthermore, the root mean squared residuals for the three panels are $0.067$, $0.081$, $0.141$, respectively. These are all larger than the corresponding values under the AIC-1 model identified in Section \ref{sec:preliminary_selection} below, which are $0.053$, $0.062$, $0.124$, respectively. The  dashed blue lines in the bottom row show LOESS fits to the AIC-1 model residuals (which are not plotted), and the middle row shows the fit of the AIC-1 model to the data.}  

\subsection{Results of applying Algorithm 3}
\label{sec:preliminary_selection}

Table \ref{tab:activity_model_results} summarizes the short-list of models obtained in Step 4 of Algorithm 3, and also the {\edits saturated and preliminary models}. The second and third columns list the number of parameters in the short-listed model and the {\edits average deviance across the 10 screening datasets input into Algorithm 3, respectively.} Columns four through six list the ranking of the models by the three {\edits selection} criteria. Columns seven through nine give the criteria values, 
$\overline{\Delta\text{AIC}}$, $\overline{\Delta\text{BIC}}$, and $\overline{\Delta\text{CV}}$ which are given by \eqref{eqn:average_crit}. 
Smaller values are preferable for all quantities listed in the table. 
{\edits Most of the top five  $\overline{\Delta\text{BIC}}$ and  $\overline{\Delta\text{CV}}$ ranked models are the same as the top $\overline{\Delta\text{AIC}}$ models, and we have only listed  each unique model once. For brevity, we denoted the top $\overline{\Delta\text{AIC}}$ ranked model by AIC-1, and use similar notation for other ranked models. }

 \begin{table}[t]
\caption{Summary of results for stage one of our model selection procedure presented in Section \ref{sec:model_selection_procedure}. The preliminary model and the top five models under the {\edits $\overline{\Delta\text{AIC}}$, $\overline{\Delta\text{BIC}}$, and $\overline{\Delta\text{CV}}$} criteria are listed. {\edits Repeated models are omitted.} 
\label{tab:activity_model_results}}
\centering
\bgroup
\def\arraystretch{1}%
\resizebox{\textwidth}{!}{
\begin{tabular}{l|rr|rrr|rrr}
  \hline
\vspace{-0.25cm} & & & & & & & &\\ 
Model &  no. paras & $\overline{\text{dev.}}$ & $\overline{\Delta\text{AIC}}$ rank & $\overline{\Delta\text{BIC}}$ rank & $\overline{\Delta\text{CV}}$ rank & $\overline{\Delta\text{AIC}}$& $\overline{\Delta\text{BIC}}$ & $\overline{\Delta\text{CV}}$ \\ 
  \hline
Saturated & 21 & 0 & 48 & 217 & 137 & 9.2 & 29.5 & 0.46\\
Preliminary & 11 & 70 & 1008 & 933 & 788 & 58.9 & 59.6 & 1.44 \\ \hline
AIC-1 &  11 & 11 & 1 & 2 & 1 & 0.0 & 0.7 & 0.00 \\ 
AIC-2 & 12 & 10 & 2 & 5 & 2 & 1.3 & 4.5 & 0.06 \\ 
AIC-3 & 11 & 13 & 3 & 3 & 5 & 1.7 & 2.4 & 0.11 \\ 
AIC-4 & 10 & 15 & 4 & 1 & 3 & 1.9 & 0.0 & 0.07 \\ 
AIC-5 &13 & 10 & 5 & 8 & 7 & 2.5 & 8.4 & 0.13 \\ 
BIC-4& 11 & 15 & 6 & 4 & 8 & 3.3 & 4.0 & 0.13 \\ 
CV-4 & 15 & 12 & 37 & 54 & 4 & 8.3 & 19.4 & 0.11 \\
   \hline
\end{tabular}}
\egroup
\end{table}

\begin{table}[t]
\caption{Maximum likelihood estimates of the AIC-1 stellar activity model coefficients for the screening dataset with the largest spot. Blank entries mean the coefficients are set to zero. All the outputs were normalized, but for interpretability the $u(t)$ coefficient estimates in m/s are given in parentheses.\label{tab:bic_fit}}
\centering
\begin{tabular}{c|cccc}
& $X$ coeff ($a_{j1}$)& $\dot{X}$ coeff ($a_{j2}$)& $\ddot{X}$ coeff ($a_{j3}$)& $Z_j$ coeff ($a_{j4}$)\\
  \hline 
$u(t)$ (m/s) &  -0.01 (-0.02) & -0.31 (-0.48) & &  \\
$q_{\text{PC}1}(t)$ &   0.50 &  &  0.08  & \\
$q_{\text{PC}2}(t)$ &  &0.31 &   &\\
   \hline
\end{tabular}
\end{table}

\begin{figure}[t]
\centering
\begin{subfigure}{0.39\textwidth}
 \centering
\includegraphics[width=1\linewidth,trim=0mm -15mm 0mm 0mm,clip]{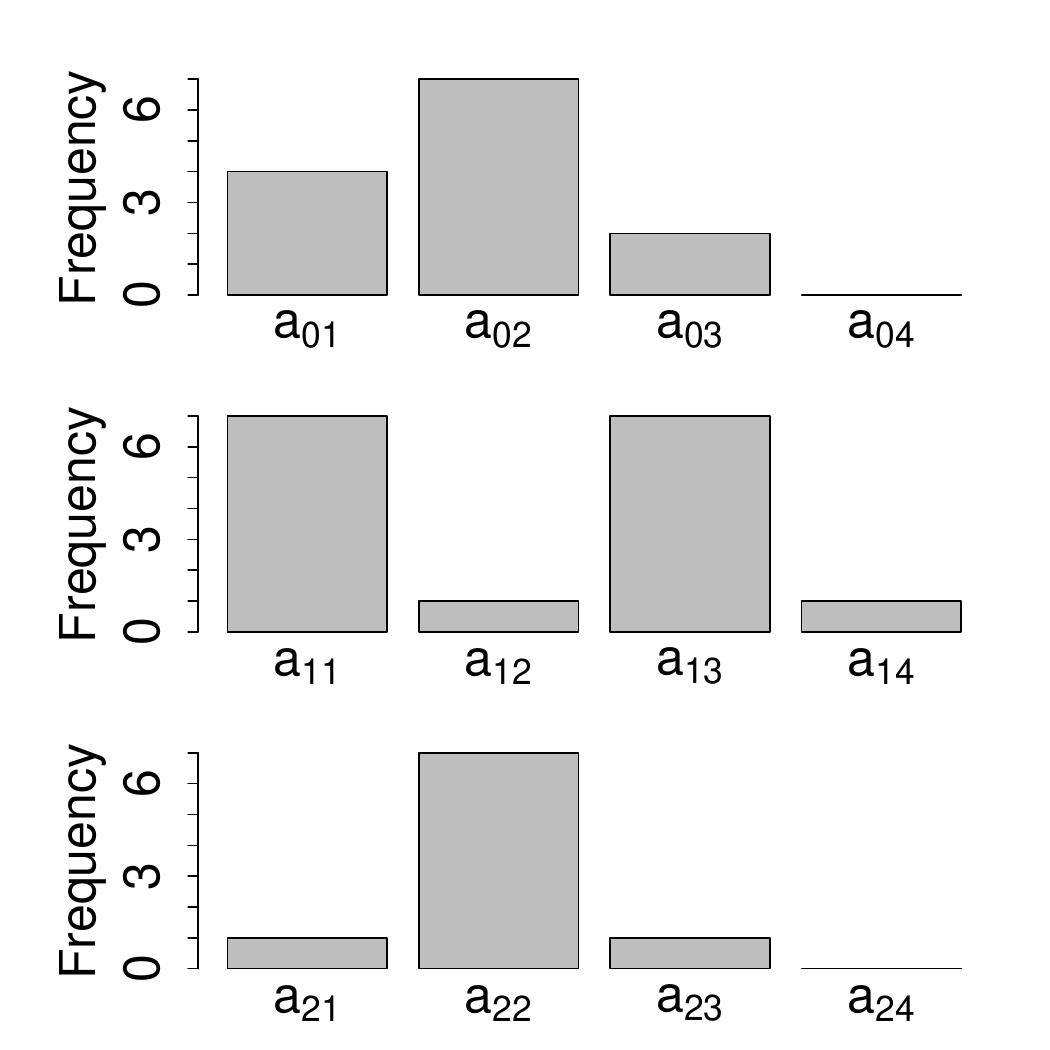}
\end{subfigure}
 \begin{subfigure}{0.6\textwidth}
   \centering
   \includegraphics[width=1\textwidth,trim=0mm 5mm 0mm 20mm,clip]{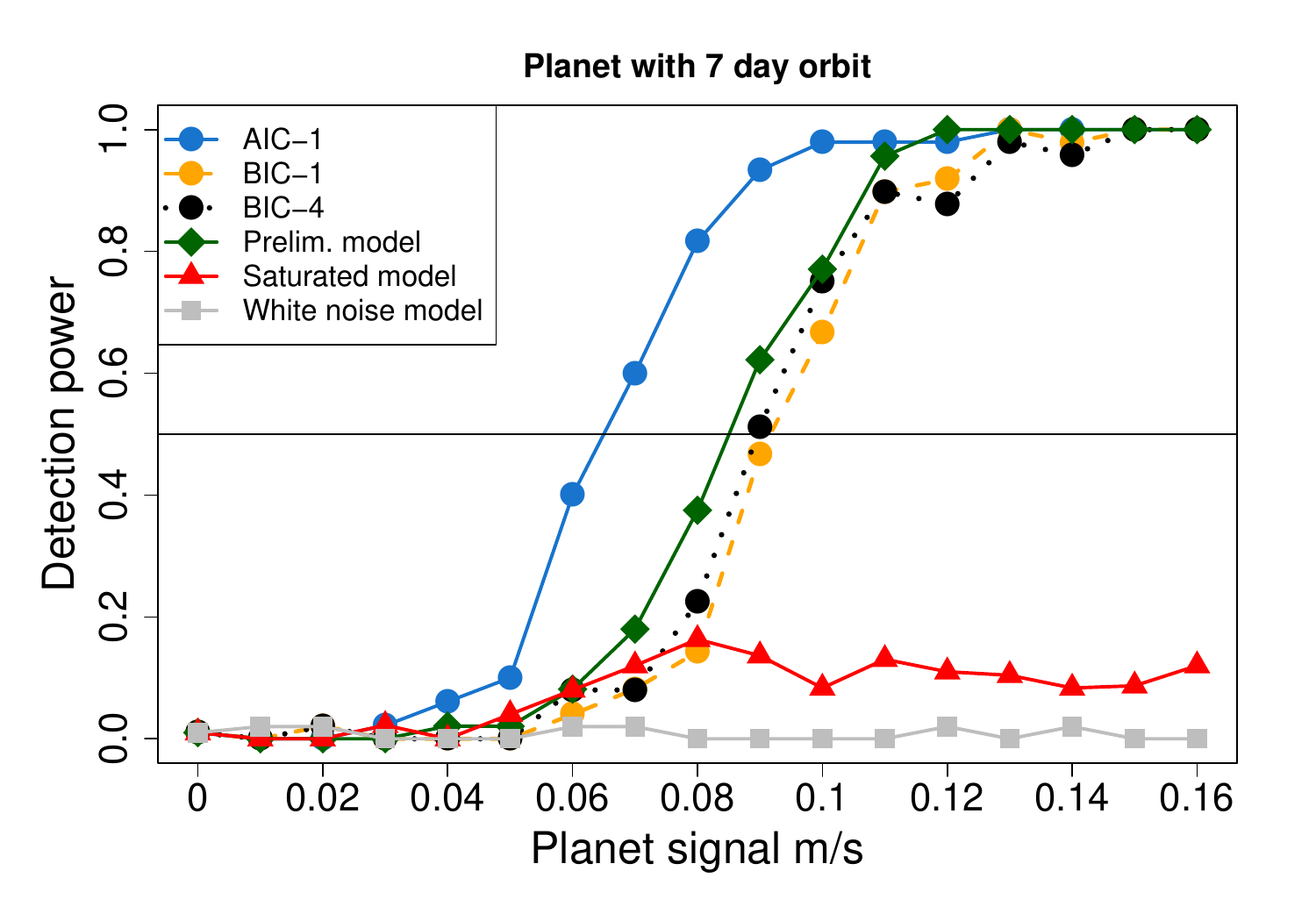}
 \end{subfigure} 
\vspace{-0.5cm}
\caption{(Left) parameter frequencies across the top {\edits ranked $\overline{\Delta\text{AIC}}$ /  $\overline{\Delta\text{BIC}}$ / $\overline{\Delta\text{CV}}$} models {\edits given} in Table \ref{tab:activity_model_results}. (Right) planet detection power for a selection of the models summarized in Table \ref{tab:activity_model_results} (left). The other models in Table \ref{tab:activity_model_results} perform similarly to the AIC-1 model.  \label{fig:rajpaul_power_tau7}}
\end{figure}

The preliminary model performs poorly under all three of the criteria in Table \ref{tab:activity_model_results}, as might be expected from the discussion in Section \ref{sec:preliminary} and the residuals shown in the bottom row of Figure \ref{fig:model_fits}. 
Table \ref{tab:bic_fit} gives the maximum likelihood estimates (MLEs) of the model coefficients for AIC-1. {\edits The MLEs of the kernel parameters are $\log \hat{\tau}_s=2.30$, $\log \hat{\lambda}_p=-0.95$, and $\log \hat{\lambda}_e= 9.07$}. {\edits These kernel parameter values are consistent} 
with a star with a single spot and a 10 day rotation period. {\edits The BIC-1 model (AIC-4 and CV-3) has non-zero coefficients $a_{02}$, $a_{11}$, $a_{13}$, and $a_{22}$, and turns out to be contained in all seven models listed in the bottom section of Table \ref{tab:activity_model_results}, i.e., these coefficients are non-zero in these seven models. This can be seen from the left panel of Figure \ref{fig:rajpaul_power_tau7}, which shows the relative frequencies of the model coefficients across the  models listed in the bottom section of Table \ref{tab:activity_model_results}.} 


Next, the right panel of Figure \ref{fig:rajpaul_power_tau7} shows a selection of the results of Step 5 of Algorithm 3, i.e., it shows the power for detecting planets under different models in Table \ref{tab:activity_model_results} at a range of planet signal amplitudes. We set the Type I error rate of the LRT to $0.01$ and thus for planet RV signals of $0$ m/s the detection power is $0.01$.  The steep slopes of the power curves is a typical feature in planet detection studies because planets and stellar activity affect observed spectra quite differently and consequently small increases in the planet RV signal amplitude can make it significantly easier to detect the planet.  In the plot, we see {\edits the AIC-1 model (blue circles, solid line) has a detection threshold of between $0.06$ m/s and $0.07$ m/s (i.e., the amplitude at which $0.5$ detection power is achieved), and  has detection power close to 1 for planet signal amplitudes $\geq 0.1$ m/s.} {\edits Four of the other top ranked models in Table \ref{tab:activity_model_results} had almost identical performance and are not plotted. The BIC-1   (AIC-4 and CV-3) model has lower power (orange circles, dashed line) because it is too simple, 
and in particular it does not include the coefficient $a_{01}$, which is small when fitted, but still impacts the deviance and detection power. The BIC-4 model (black circles, dotted line) is the same except it includes the coefficient $a_{21}$, which seems to be  superfluous in that it does not lower the average deviance noticeably,  and indeed BIC-4 performs similarly to BIC-1. }
In summary, several models perform almost identically well, but we set $\mathcal{M}_{\text{\text{MS-G2V}}}$ to be the AIC-1 model because it has the best overall rank (the sum of the three slection criteria ranks).

The preliminary model (green diamonds) performs substantially worse  than the AIC-1 model, but slightly better than the BIC-1 and  BIC-4 models. The saturated model (red triangles) is overly flexible and therefore absorbs planet signals, which results in low power across the range of planet signal amplitudes considered.   
The white noise model (solid squares) considers only the RV signal and treats any RV corruption from stellar activity as independent Gaussian realizations with a fixed standard deviation (plus measurement error).  This approach can be valuable for analyzing legacy RV datasets that typically have sparsely spaced observations and no stellar activity information, see \citet{ford2006jitter}.  However, it  results in almost no power for the small planet signal amplitudes considered in Figure \ref{fig:rajpaul_power_tau7}. 

{\edits In addition to detecting planets it is of interest to infer their properties, and Appendix \ref{app:estimation} summarizes the orbital period ($\tau_p$) and signal amplitude ($K$) estimation results for the AIC-1 model. Another important consideration is how planet detection power varies as a function of period, and in preliminary studies we found that the power is lower for planets with orbital periods that coincide with the stellar rotation period or its harmonics.}

\subsection{Comparison to Rajpaul et al. (2015a)}
\label{sec:rajpaul_comparison}

That the preliminary model performs poorly compared to the AIC-1 model (see Figure \ref{fig:rajpaul_power_tau7})  is  unsurprising because the model was originally designed  by \citetalias{rajpaul2015gaussian} to capture the evolution of the RV corruption, normalized flux (or $\logrhk$), and $\bis$, rather than the evolution of our  indicators.  
Therefore, we repeat our detection power analysis for the \citetalias{rajpaul2015gaussian} model, but using the \citetalias{rajpaul2015gaussian} activity indicators in order to compare their utility for planet detection to that of our PCA based activity indicators. 
Both normalized flux and BIS were computed {\edits during the SOAP 2.0 runs used to create the $Y_A$ datasets, and we add noise consistent with that of the  $Y_A$ datasets}. 
{\edits Specifically, for} normalized flux we computed the noise levels based on a Poisson model for variations in the number of photons detected. 
For BIS, we assumed a constant standard deviation of 0.0198 m/s, where the specific value was motivated by scaling actual BIS error measurements in \citet{dumusque2012earth} to account for the increased spectral resolution and SNR of our input data.

\begin{figure}[t]
\centering
\begin{subfigure}{.6\textwidth}
  \centering
   \includegraphics[width=1\linewidth,trim=0mm 5mm 0mm 20mm,clip]{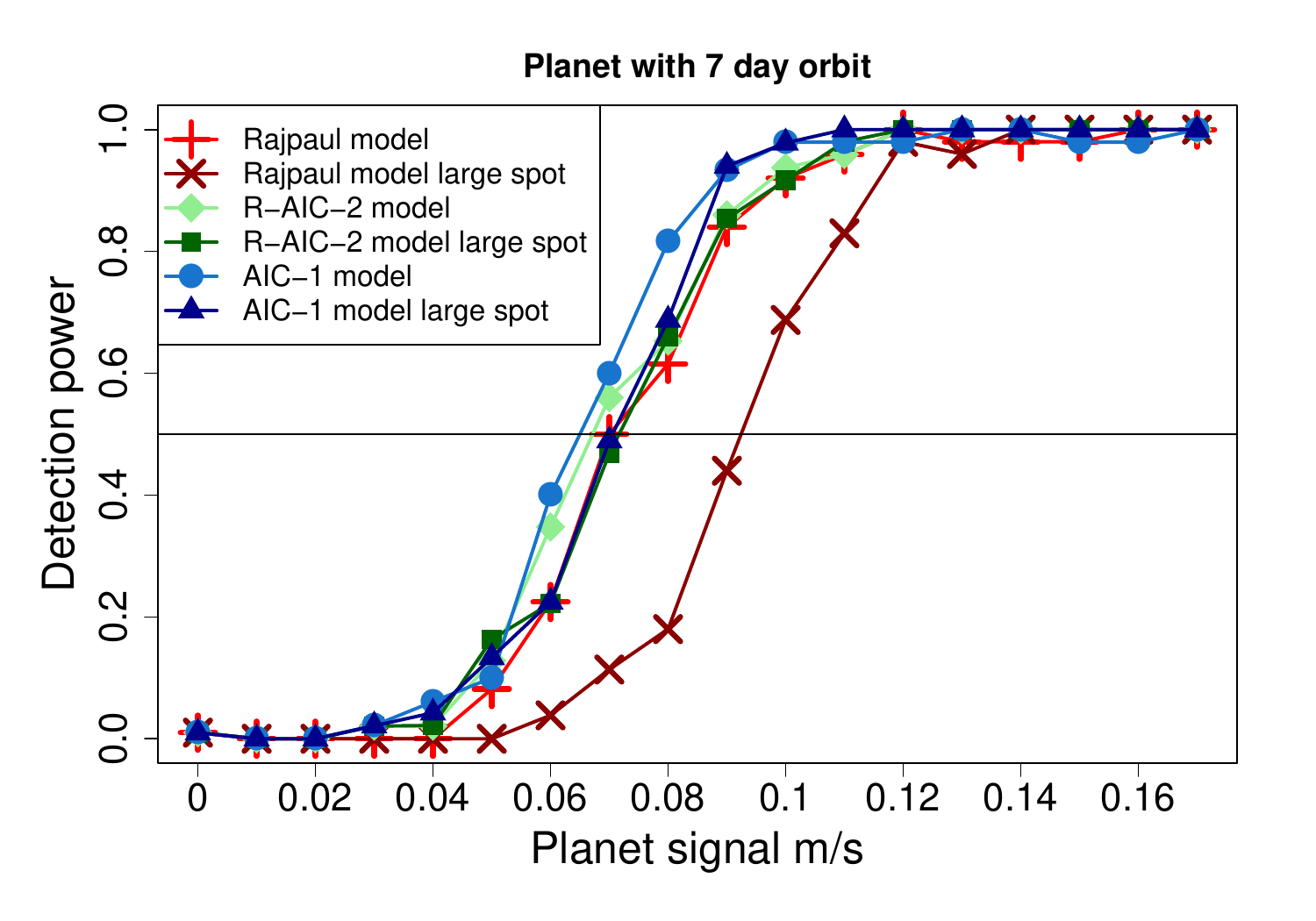}
\end{subfigure}%
\begin{subfigure}{.39\textwidth}
  \centering
  \includegraphics[width=1\linewidth,trim=0mm 10mm 0mm 20mm,clip]{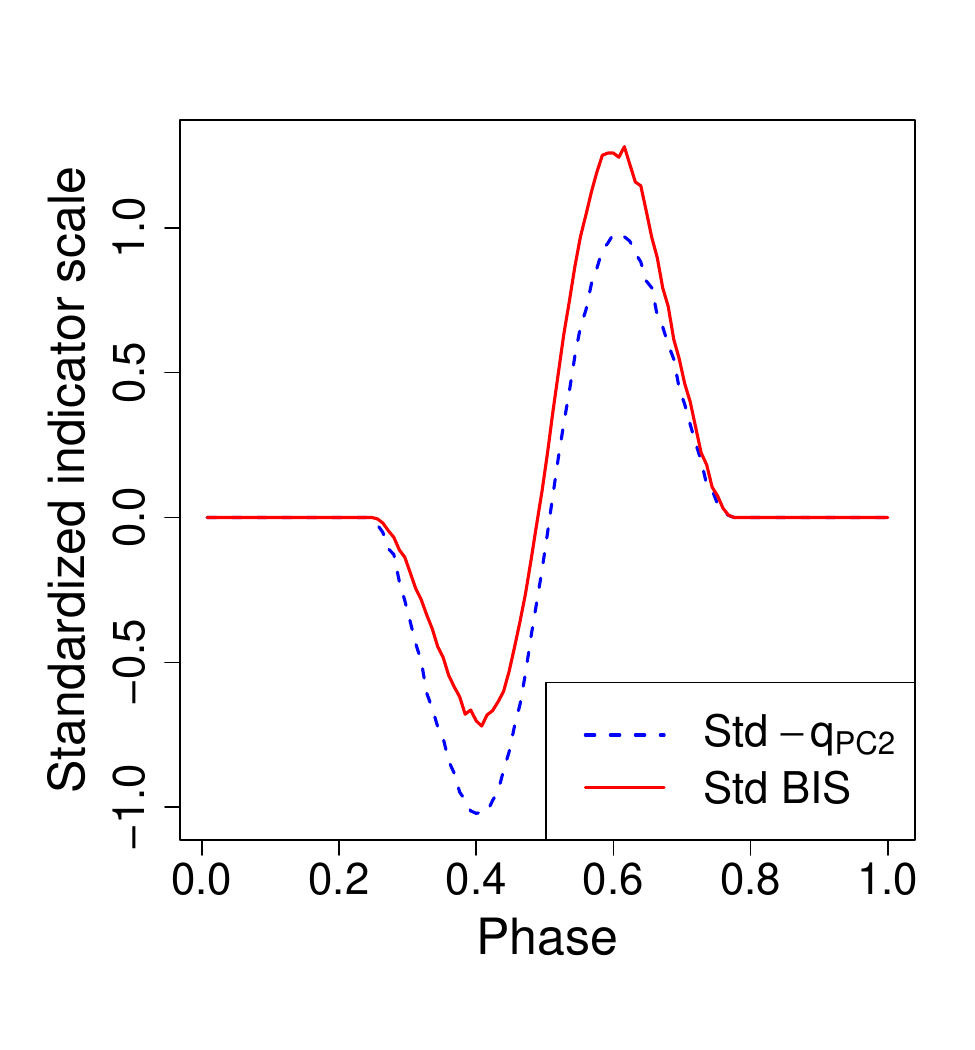}
\end{subfigure}
\caption{The left panel shows the detection power under the \citetalias{rajpaul2015gaussian} model  using their normalized flux and BIS indicators (red `+' symbols), and again in the case where the datasets with planets all have a large spot (dark red `$\times$' symbols). Also shown is the detection power for the R-AIC-2 model (green diamonds), and again in the large spot case (dark green squares). For comparison, the AIC-1 model results (with our proxies) is reproduced from Figure \ref{fig:rajpaul_power_tau7} (blue circles), and the AIC-1 model detection power in the case of a large spot is plotted (dark blue triangles). The right panel compares the $q_{\text{PC}2}$ and BIS indicators computed from the  noiseless data $Y_s$. The indicators are standardized (centered and normalized) for the purpose of the comparison. \label{fig:rajpaul_power}}
\end{figure}

The red ``+" symbols in the left panel of Figure \ref{fig:rajpaul_power} show the detection power under the \citetalias{rajpaul2015gaussian} model when using the normalized flux and BIS  activity indicators described above (for a planet with a 7 day orbital period). 
For comparison, the detection power under the AIC-1 (blue circles) is re-ploted from Figure \ref{fig:rajpaul_power_tau7}). The \citetalias{rajpaul2015gaussian} model has a detection threshold of about 0.07 m/s and therefore performs almost as well as the AIC-1 model, and better than the preliminary model (see Figure \ref{fig:rajpaul_power_tau7}). {\edits However, the dark red ``$\times$" symbols show that the detection power of the \citetalias{rajpaul2015gaussian} model is substantially reduced when detecting planets in the presence of an 866 micro hemispheres (MSH) stellar spot, which is the 95\% quantile of the spot size distribution \eqref{eqn:size} used in our main simulations. In other words, in the case of particularly active MS-G2V stars, the \citetalias{rajpaul2015gaussian} model has reduced performance. In contrast, the dark blue triangles show that the detection power of the AIC-1 model is almost unaffected by the presence of such a large spot. }

It is unclear if the superiority of the AIC-1 model in the case of large spots is due to  our PCA based indicators or our model selection procedure or both. To investigate this we applied our model selection procedure in the case of the normalized flux and BIS indicators. We again found the five best $\overline{\Delta\text{AIC}}$, $\overline{\Delta\text{BIC}}$, and  $\overline{\Delta\text{CV}}$ ranked models. The  top performing model in terms of  detection threshold was that ranked second by $\overline{\Delta\text{AIC}}$, which we denote  by R-AIC-2. Figure \ref{fig:rajpaul_power_tau7} shows the detection power of the R-AIC-2 model (green diamonds) and its detection power in the case of a large spot (dark green squares). The R-AIC-2 model performs similarly to the AIC-1 model in both cases, and substantially better than the  \citetalias{rajpaul2015gaussian} model  in the case of large spots. The R-AIC-2 model and the maximum likelihood estimates of its coefficients are given in Table \ref{tab:rcv3_fit} in Appendix \ref{app:rcv3_paras}.  

In conclusion, 
for the simple case of a single constant spot it is principally our model selection procedure rather than our PCA based stellar activity indicators that offers improved detection power, and this improvement is mainly obtained in the case of large spots, as might be expected. Therefore, for the time being, astronomers may  prefer to keep using normalized flux (or $\logrhk$) and BIS as indicators, but the model should be updated to the R-AIC-2 model or a similar model. In particular, although the \citetalias{rajpaul2015gaussian} model has the appealing feature of being physically motivated, it seems that planet detection power can be improved by using a more flexible model that can better capture the stellar activity time series jointly, at least in the case of a constant spot. 
For cases where the stellar spectra are available to the investigators, it may be preferable to use our PCA based indicators for computational and robustness reasons, because the AIC-1 model (for our indicators) requires five fewer parameters than the R-AIC-2 model (for the \citetalias{rajpaul2015gaussian} indicators). Alternatively, investigators may wish to transform the BIS indicator so that the blueshifts (troughs) and redshifts (peaks) are more symmetric because this will likely make it possible to use a simpler GP model. 
The right panel of Figure \ref{fig:rajpaul_power} compares $-q_{\text{PC}2}$ and BIS (after centering) for the noiseless data $Y$. The asymmetry suggests that scaling blueshifts and redshifts differently could be a good choice of transformation.

With a longer term perspective, we emphasize that our indicators were constructed automatically without the need for physical derivations.  This is important because it means that our approach can likely be easily generalized to more complex stellar activity phenomena, such as evolving spots, where different or additional indicators may be needed. The PCA approach ensures that when additional indicators are used they contain different information to those already included, whereas there is no such guarantee for expert-identified indicators chosen for their individual interpretations.  Similarly, our indicator construction method could potentially also allow custom indicators to be used for each star or type of star, and this is discussed further in Section \ref{sec:generalizing}. 

\section{Discussion}
\label{sec:discussion}


\subsection{Scope of our framework}

Our general procedures apply broadly, and are not limited to the specific indicator and model choices discussed here, as we now explain.  Firstly, in Algorithm 1, PCA can be straightforwardly replaced by another dimension reduction technique.  
We are currently exploring the use of diffusion maps \citep[e.g.,][]{coifman2005geometric,coifman2006diffusion} which do not have the constraint of projecting onto a linear subspace as does PCA. Capturing non-linearity  could be particularly helpful for more complex situations, such as when stellar spots evolve over time.  Our initial investigations also suggest that capturing non-linear structure may  help to improve planet detection power when there is  aliasing between the stellar rotation period and the planet orbital period. 

Secondly, our automated  model selection procedure allows investigators to easily find the best model for new stellar variability indicators, and to compare 
the relative performance of competing 
indicators. Lastly, although we have assumed a single spot of constant (but unknown) size, our ongoing work suggests that our class of models is able to capture evolving spots. More generally, the extent to which it is reasonable to use the specific class of models (\ref{eqn:model1})-(\ref{eqn:our_errors}) for modeling complex stellar activity is a topic of future work. 

\subsection{Generalizing to other star types}\label{sec:generalizing}

A key direction for expanding the scope of our framework is to generalize to other star types, and there are two main aspects to this task, as we now discuss. Our Doppler-constrained stellar activity indicators defined by $\mathcal{I}_{\text{\text{MS-G2V}}}$ are obtained by applying Algorithm 1 to $Y_s$, which contains noiseless spectra of an MS-G2V star, the Sun. For stars unlike MS-G2V stars other indicators may be better, but we do not have noiseless datasets from other types of star to which we can apply Algorithm 1 to derive new indicators. 
One possible solution is to apply  Algorithm 1 to every dataset and thereby obtain star-specific indicators from the noisy spectra observed. We now briefly investigate the potential of this approach. 

\begin{figure}[t]
 \centering
 \includegraphics[width=0.38\linewidth,trim=0mm 0mm 0mm 0mm,clip]{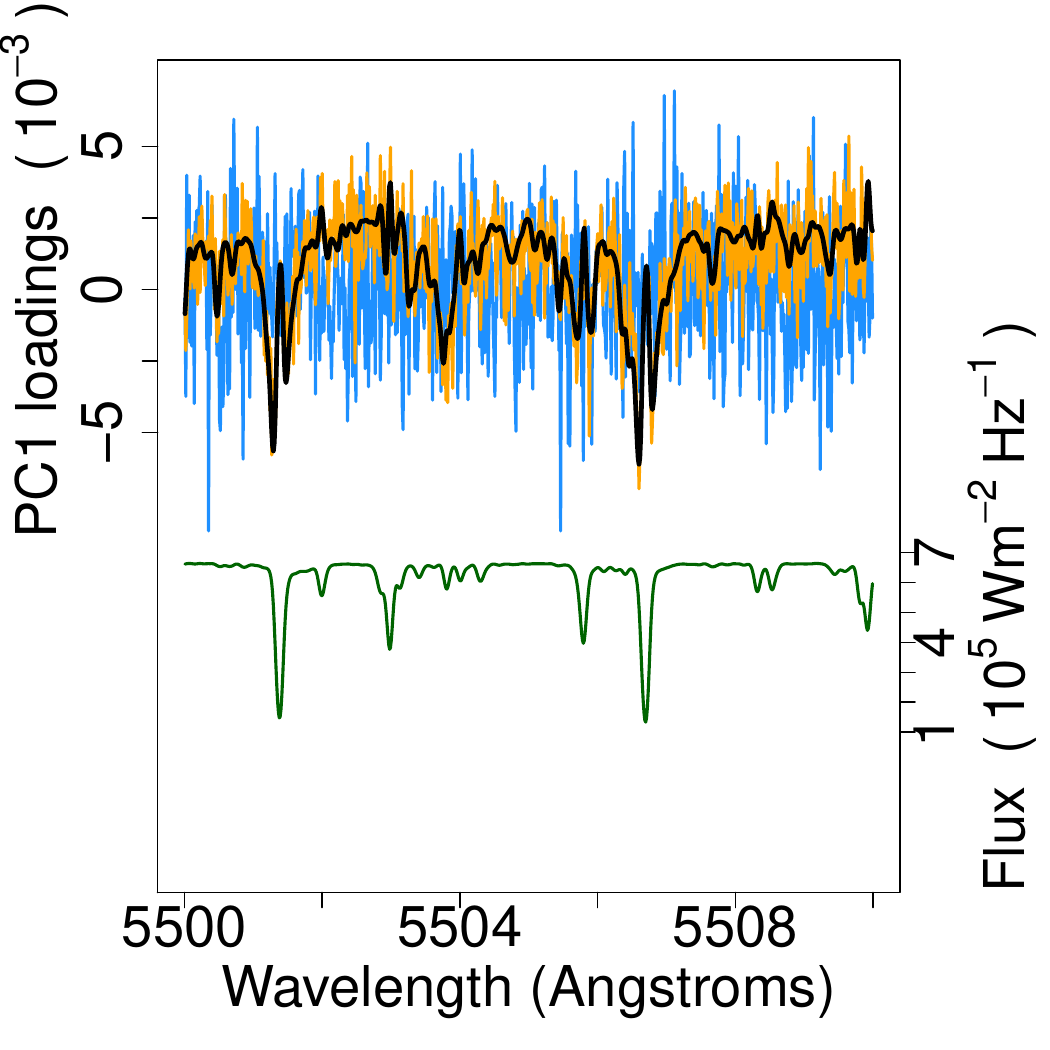}
\includegraphics[width=0.60\linewidth,trim=0mm 5mm 0mm 20mm,clip]{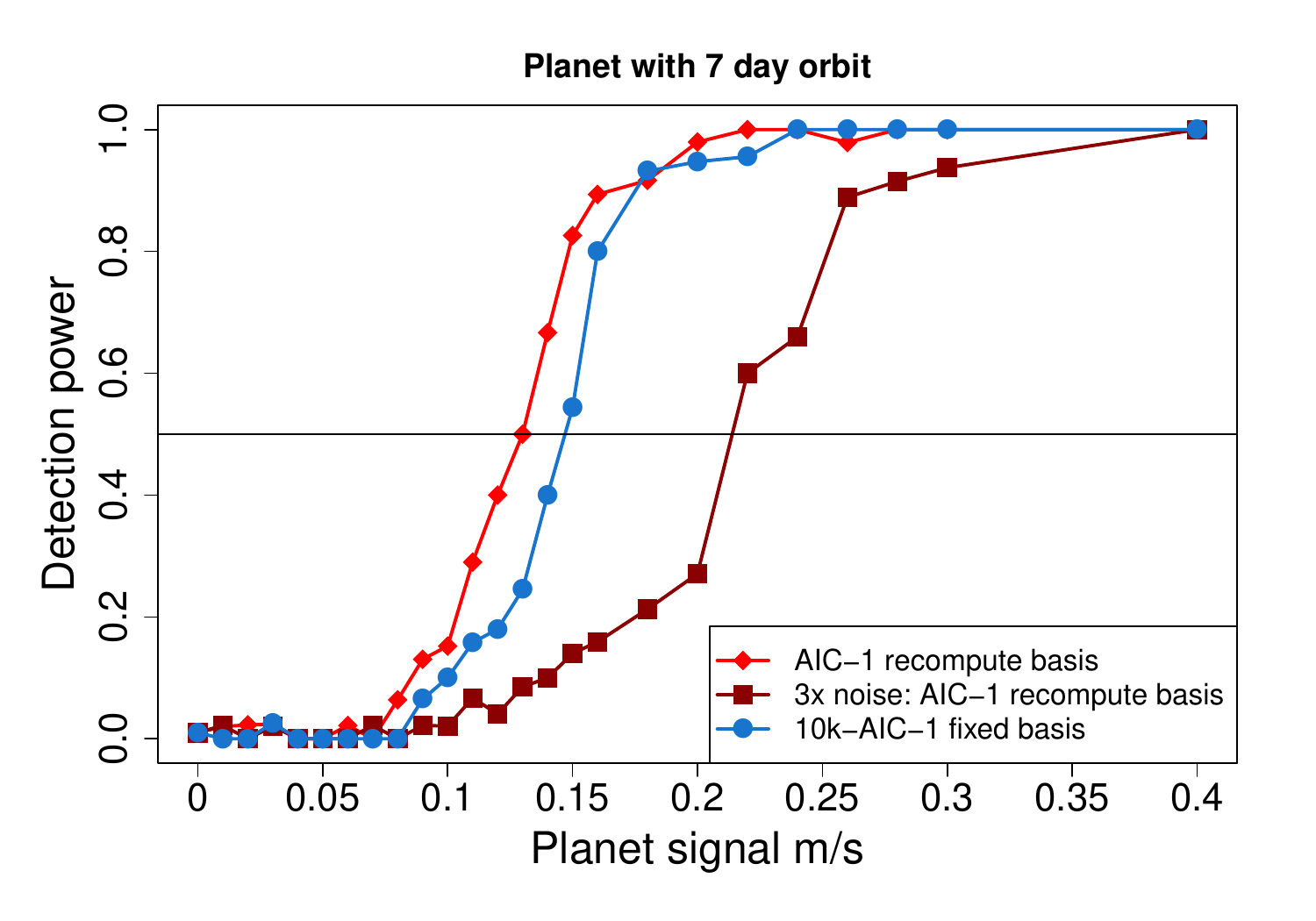}
\vspace{-0.0cm}
\caption{(Left) PC1 loadings (for a small portion of the input wavelength range) obtained by applying Algorithm 1 to a noisy dataset with a 661 MSH spot (blue line, upper half) and a 10k MSH spot (orange green line, upper half). For comparison, the corresponding loadings from $\mathcal{I}_{\text{\text{MS-G2V}}}$ are also shown (thick black line, upper half). In the bottom half of the panel a small segment of the quiet spectrum is plotted (green).  (Right) planet detection power under the AIC-1 model when the activity indicator vectors are recomputed for each noisy dataset (red diamonds), in the case where all null and test simulations use a 10k MSH spot. For comparison we also plot the detection power when the noise standard deviation is inflated by a factor of three (dark red squares), i.e., SNR$=167$, and when we use the fixed indicators $\mathcal{I}_{\text{\text{MS-G2V}}}$ with the 10k-AIC-1 model (blue circles). \label{fig:noisy_power}}
\end{figure}

The thin blue line in the upper part of the left panel of Figure \ref{fig:noisy_power} shows a section of the principal component corresponding to $q_{\text{PC1}}$ when Algorithm 1 is applied directly to a noisy dataset containing a 661 MSH spot (fairly large under the spot size distribution \eqref{eqn:size}). For comparison, the think black line  in the upper part of the left panel shows the corresponding principal component contained in $\mathcal{I}_{\text{\text{MS-G2V}}}$ (computed from noiseless data). The bottom part of the plot shows the corresponding section of the spectrum (green). It is clear that in the case of the noisy dataset, the principal component is dominated by noise, and indeed we found it almost useless for planet detection. Here, we instead focus on the case of a very large 10k MSH spot, i.e., in both null and test simulations in Algorithm 2 we fix the spot size at 10k MSH, rather than simulating it from \eqref{eqn:size}. The thin orange line in the upper part of the left panel of Figure \ref{fig:noisy_power} shows that in this case the first indicator in $\mathcal{I}_{\text{\text{MS-G2V}}}$ can be reasonably approximated from one of the noisy datasets.  The right panel shows the resulting detection power under the AIC-1 model (see Table \ref{tab:bic_fit}), for this 10k MSH spot case, when recomputing  the indicator vectors for each dataset by applying Algorithm 1 (red diamonds). In this large spot scenario, although  the AIC-1 model was found to still be the best model if  recomputing  the indicator vectors for each dataset, this was not the case when using the fixed indicators defined by $\mathcal{I}_{\text{\text{MS-G2V}}}$. Thus, for comparison with the fixed indicator case, 
we plot the new top ranked $\overline{\Delta\text{AIC}}$ model, denoted 10k-AIC-1 (blue circles) and given in Appendix \ref{app:10kaic1_paras}.\footnote{Interestingly, in our preliminary studies with lower SNR, the AIC-1 model was still the best model in the large spot case, suggesting that SNR is an important factor in the model selection.}  In conclusion, there appears to be no loss of power due to re-computing the indicators vectors, in the case of a 10k MSH spot, and in fact there is a small gain in power. Of course, even for large spots, if the spectra are sufficiently noisy then we expect that the detection power will decrease, especially when recomputing the indicators for each dataset. Confirming this, the dark red squares  show the detection power when the noise is inflated by a factor of three (i.e., SNR$=167$), and we recompute the indicators for each dataset. 

The left panel of Figure \ref{fig:noisy_power} suggests that it may be fruitful to apply a penalization or smoothing approach when deriving stellar activity indicators from noisy datasets. It also suggests that observing stars during high activity periods could be useful for deriving reliable indicators, whereas in the past observing currently highly active stars has been intentionally avoided (in exoplanet surveys).   

{\edits The second key challenge is that for star types other than MS-G2V the distribution of spot related activity is less well understood, and may not be well approximated by the distribution for the Sun (as is done in Algorithm 2). A potential solution is to  use  a hierarchical Bayesian structure to jointly model stellar activity from multiple stars with similar spectroscopic properties. This would allow us to infer population-level stellar activity parameters and a corresponding model that could be used for generating replicate datasets from such stars. 
The population inference could also be used to infer indicator vectors (e.g., $\mathcal{I}_{\text{\text{MS-G2V}}}$), thereby avoiding application of Doppler-constrained PCA directly to the noisy spectra of individual stars. }

{\edits \subsection{Parameters}

}


\vspace{-0.5cm}
\section*{Acknowledgements}
The authors thank two anonymous reviewers, an anonymous Associate Editor, and Editors Nicoleta Serban and Tilmann Gneiting for constructive comments and references which helped improve the paper. The authors also thank Nathan Hara for helpful comments.
This material was based upon work partially supported by the National Science Foundation under Grant DMS-1127914 to the Statistical and Applied Mathematical Sciences Institute (SAMSI). Any opinions, findings, and conclusions or recommendations expressed in this material are those of the authors and do not necessarily reflect the views of the National Science Foundation.
E.B.F. acknowledges support from the Penn State Eberly College of Science and Department of Astronomy \& Astrophysics, the Center for Exoplanets and Habitable Worlds, and the Center for Astrostatistics.  This work was funded in part by NSF AST award \#161086 and NASA Exoplanets Research Program \#NNX15AE21G.

\bibliographystyle{imsart-nameyear}
\bibliography{improving_planet_detection_power.bib}

\newpage
\begin{appendices}



\section{Doppler component for Algorithm 1}
\label{app:dopplerw}

Let the star light intensity  at rest frame wavelength $\lambda^{(0)}$, when normalized by the total light intensity across wavelengths, be denoted by $f(\lambda^{(0)})$, so that the function $f(\cdot)$ gives the rest frame  normalized stellar spectrum (at some fixed time).  For a source moving with a radial velocity of $v=cz$ the Doppler shift is $z$, where $c$ is the speed of light in m/s, i.e., the observed intensity at observed wavelength $\lambda$ is given by $f(\lambda/(1+z))$. Thus, a Taylor expansion tells us that the observed intensity at observed wavelength $\lambda$ will approximately be given by $f(\lambda) - z f'(\lambda)$, where $f'=df/d\log\lambda = \lambda df/d\lambda$. Since the relevant Doppler shifts are typically very small ($z\approx 10^{-8}$) this Taylor approximation is very accurate and we can therefore represent Doppler shifts as scalar multiples of the vector $(f'(\lambda_1),\dots,f'(\lambda_p))^T$, where $\lambda_i$ is the recorded wavelength, for $i=1,\dots,p$. 

A complication is that $f'(\cdot)$ is not known exactly, and the derivative of observed stellar spectrum $f'_t(\cdot)$ changes over time due to the presence of stellar activity, any planets, and noise, see \eqref{eqn:redshift}. 
However, fortunately, the temporal changes to $f_t'(\cdot)$ can be regarded as second-order, so it is reasonable to treat $\boldsymbol{w}$ as fixed across time. Here, we compute the mean observed spectrum across the different observation times, denoted $\bar{f}(\cdot)$, and set $\boldsymbol{w}=(\bar{f}'(\lambda_1),\dots,\bar{f}'(\lambda_p))^T$. 

\section{Kepler planet model details}\label{app:planet}

The true anomaly function $\phi(t)$ in (\ref{eqn:planet}) is given by solving the following system of equations
\begin{align}
\tan \frac{\phi(t)}{2} &= \left(\frac{1+e}{1-e}\right)^{\frac{1}{2}} \tan \frac{E(t)}{2}\\
E(t)-e \sin E(t) &= M(t)\\
M(t) &= \frac{2 \pi t}{\tau_p} + M_0,
\end{align}
where $\tau_p$ is the orbital period of the planet, $M_0$ is known as the mean anomaly at $t=0$, and $e$ is the orbital eccentricity. 

\section{Covariance matrix calculation}\label{app:C}

Here we specify the covariance matrix $\Sigma$ implied by the model (\ref{eqn:model1})-(\ref{eqn:model_class}) and the covariance function  (\ref{eqn:kernel}). Let $A^{(a,b)}$ be the matrix whose $(i,i')$ entry is $\text{Cov}\left(\frac{d^a}{dt_i^a}X(t_i),\frac{d^b}{dt_{i'}^b}X(t_{i'})\right)$, for $a,b=0,1,2$. Then, for model parameters $a_{jk}$, for $j=0,\dots,\indnum$ and $k=1,2,3,4$, the $n\times n$ diagonal block of $\Sigma$ corresponding to output $j$ (i.e., the square block with rows and columns $jn+1,\dots,(j+1)n$) is given by 
\begin{align}
\sum_{k_1=1}^3\sum_{k_2=1}^3 a_{jk_1}a_{jk_2}A^{(k_1-1,k_2-1)}
 + a_4^2B^{(0,0)},
\end{align}
for $j=0,\dots,\indnum$, where the $(i,i')$ entry of $B^{(0,0)}$ is $\text{Cov}(Z_0(t_i),Z_0(t_{i'}))$ (recall $Z_0, \dots,Z_\indnum$ all have the same covariance function parameters $\phi_Z$). Similarly, the off-diagonal $n\times n$ block corresponding to the covariance between outputs $j_1$ and $j_2$ (i.e., the square block with rows  $j_1n+1,\dots,(j_1+1)n$ and columns $j_2n+1,\dots,(j_2+1)n$) is given by 
\begin{align}
\sum_{k_1=1}^3\sum_{k_2=1}^3 a_{j_1k_1}a_{j_2k_2}A^{(k_1-1,k_2-1)}.
\end{align}
Thus, all that remains is to specify $A^{(a,b)}_{ii'}$, for $a,b=0,1,2$. The term $A_{ii'}^{(0,0)}$ is (\ref{eqn:kernel}) with $t=t_i$ and $t'=t_{i'}$, and the remaining terms are given by
\begin{align}
A_{ii'}^{(0,1)}&=-A_{ii'}^{(1,0)}=T_1A_{ii'}^{(0,0)}\\
A_{ii'}^{(1,1)}&=-T_1A_{ii'}^{(0,1)} + T_2A_{ii'}^{(0,0)}\\
A_{ii'}^{(0,2)}&=A_{ii'}^{(2,0)}=-A_{ii'}^{(1,1)}
\end{align}
\begin{align}
A_{ii'}^{(1,2)}&=-A_{ii'}^{(2,1)}=T_1A_{ii'}^{(1,1)} + 2T_2A_{ii'}^{(0,1)} + 2T_3A_{ii'}^{(0,0)}\\
A_{ii'}^{(2,2)}&=
-T_1A_{ii'}^{(1,2)}+ 3T_2A_{ii'}^{(0,2)} - 6T_3A_{ii'}^{(0,1)} + 4T_4A_{ii'}^{(0,0)}
\end{align}
where, writing $\lambda_{ij}=2\pi(t_i-t_j)/\tau$, 
\begin{align}
T_1 &= \frac{\pi\sin(\lambda_{ij})}{2\tau_s\lambda_p^2} + \frac{t_i-t_j}{\lambda_e^2}\\
T_2 &= \frac{\pi^2\cos(\lambda_{ij})}{\tau_s^2\lambda_p^2} + \frac{1}{\lambda_e^2}\\
T_3 &= \frac{\pi^3\sin(\lambda_{ij})}{\tau_s^3\lambda_p^2}\\
T_4 &= \frac{\pi^4\cos(\lambda_{ij})}{\tau_s^4\lambda_p^2}.
\end{align}

\section{Cross validation details}\label{app:CV}

Here we describe the CV approach used in Algorithm 3, see Section \ref{sec:selection_practice} and Section \ref{sec:model_selection_procedure}.  We first investigated standard leave-one-out cross-validation but found that the models favored by this approach tend to be overly complex  and have low power for planet detection. This is unsurprising because  leave-one-out prediction is substantially easier than identifying the component of an apparent RV signal that is due to stellar activity. Indeed, the latter case is more similar to predicting the RV corruption for {\it all} the observations. Therefore, to somewhat better approximate the problem at hand, we instead leave out blocks of observations. 

In a single repetition of our CV procedure we randomly select a test block of $b=5$ consecutive observation times to hold back. Then we find the maximum likelihood parameter estimates $\hat{\theta}_T$ based on the observations at the remaining $n-b$ times, which we refer to as the training data. Let the subscripts  $B$ and $T$ attached to vectors or matrices denote elements corresponding to the test block and training data, respectively. In the case of matrices the first subscript refers to the rows and the second refers to the columns. Dropping constants, our cross validation score is
\begin{align}
\frac{1}{2}\left((\boldsymbol{s}_B-\boldsymbol{\hat{\mu}})^T\hat{V}^{-1}(\boldsymbol{s}_B-\boldsymbol{\hat{\mu}}) - \log\left|\hat{V}\right|\right)
\label{eqn:cvk}
\end{align}
where $\boldsymbol{\hat{\mu}}=\boldsymbol{\hat{m}}_B - \hat{\Sigma}_{BT}\hat{\Sigma}_{TT}^{-1}(\boldsymbol{s}_T-\boldsymbol{\hat{m}}_T)$ and $\hat{V} = \hat{\Sigma}_{BB} - \hat{\Sigma}_{BT}\hat{\Sigma}_{TT}^{-1}\hat{\Sigma}_{TB}$. Here $\hat{\Sigma}$ is the estimated covariance matrix constructed using all the observation times $\boldsymbol{t}$ and the estimated model parameters $\hat{\theta}_T$. Similarly, $\boldsymbol{\hat{m}}$ is the maximum likelihood estimate of $\boldsymbol{m}$ based on the training data. Thus, the cross validation score is the negative log conditional likelihood of the data held back under the parameters $\hat{\theta}_T$ and conditional on the training data. It makes sense to use the log conditional likelihood rather that the log likelihood because our Gaussian process model is non-parametric, which means the unconditional likelihood of the test block observations is not very informative about the predictive power of the model. We repeat the above cross validation procedure $B=10$ times for each model. The final cross validation score for a given model is 
\begin{align}
\text{CV}= \frac{1}{B}\sum_{k=1}^B\text{CV}_k
\label{eqn:cv}
\end{align}
where  $\text{CV}_k$  denotes the value of (\ref{eqn:cvk}) for the $k$th repetition of the procedure, for $k=1,\dots,B$. To ensure a fairer comparison of the models, we re-use the same 10 test blocks for all models.  

\section{Details of optimization procedure}\label{app:D}

Optimization of the parameters of the models in the class (\ref{eqn:model1})-(\ref{eqn:model_class}) and additional planet signal parameters was mostly straightforward, but there were three aspects of our approach that were specific to the context, and we detail them here. Firstly, the parameters were divided into four blocks: (i) the model coefficients $a_{jk}$, for $j=0,1,2$,  $k=1,2,3,4$, (ii) the covariance function parameters $\phi$ for $\gp$, (iii) the covariance function parameters $\phi_Z$ for $Z_j$, for $j=0,1,2$, and (iv) the mean function parameters including $m_j$, for $j=0,1,2$ and, in the case of the full model, the planet parameter vector $\alpha$. Block (i) was optimized, followed by block (ii), and so on. We iterated through the blocks until we had completed at least 10 cycles and the log-likelihood (\ref{eqn:likelihood}) converged. 

Secondly, rather than directly optimizing the full log-likelihood, we first optimized the log-likelihood for stellar indicator $q_{\text{PC}1}$ (again using parameter blocks as described above), and then the log-likelihood for $q_{\text{PC}1}$ and $q_{\text{PC}2}$, and finally the full log-likelihood. At each stage the optimized parameters were input into the next stage as initial values. This approach proved more successful than direct optimization of the full log-likelihood because the stellar activity indicators are unaffected by potential planet signals making it easier to find the global mode of their log-likelihood, and because in our case the indicators have a natural information ordering (for parameter estimation) in that the  measurement errors of the $q_{\text{PC}1}$ observations are smaller than those of $q_{\text{PC}2}$. We repeated the above procedure for 10 initializations of the parameters and chose the run that resulted in the highest log-likelihood. 

Thirdly, although most of the optimization was done using standard functions in the R software package,  period and angle parameters required more care. The period parameter $\tau_s$ in the covariance function (\ref{eqn:kernel}) was optimized using a fine grid search. To optimize the planet parameters we first performed a fine 2D grid search on $\tau_p$ and $M_0$, where for each candidate pair $(\tau_p,M_0)$ we used regression to quickly optimize the other planet parameters. In particular, following \citet{loredo2012}, we re-wrote (\ref{eqn:planet}) as $v(t) = \beta_0+\beta_1(e+\cos(\phi(t))) + \beta_2 \sin(\phi(t))$, where $\beta_0=\gamma=m_0$, $\beta_1=K \cos(\omega)$, and $\beta_2=-K\sin(\omega)$. The linear parameters $\beta_0$, $\beta_1$, and $\beta_3$ were then inferred by regressing the residuals for the radial velocity observations under the current stellar activity model fit against $e+\cos(\phi(t))$ and $\sin(\phi(t))$. For this step $e$ was fixed at a typical value (in practice there is little information to infer $e$). After the initial grid search, the parameters values found were used to initialize a joint gradient ascent optimization of the planet parameters.

\section{Estimation of Keplerian planet model parameters}
\label{app:estimation}

\begin{figure}
\centering
\begin{subfigure}{.49\textwidth}
  \centering
  \includegraphics[width=1\linewidth]{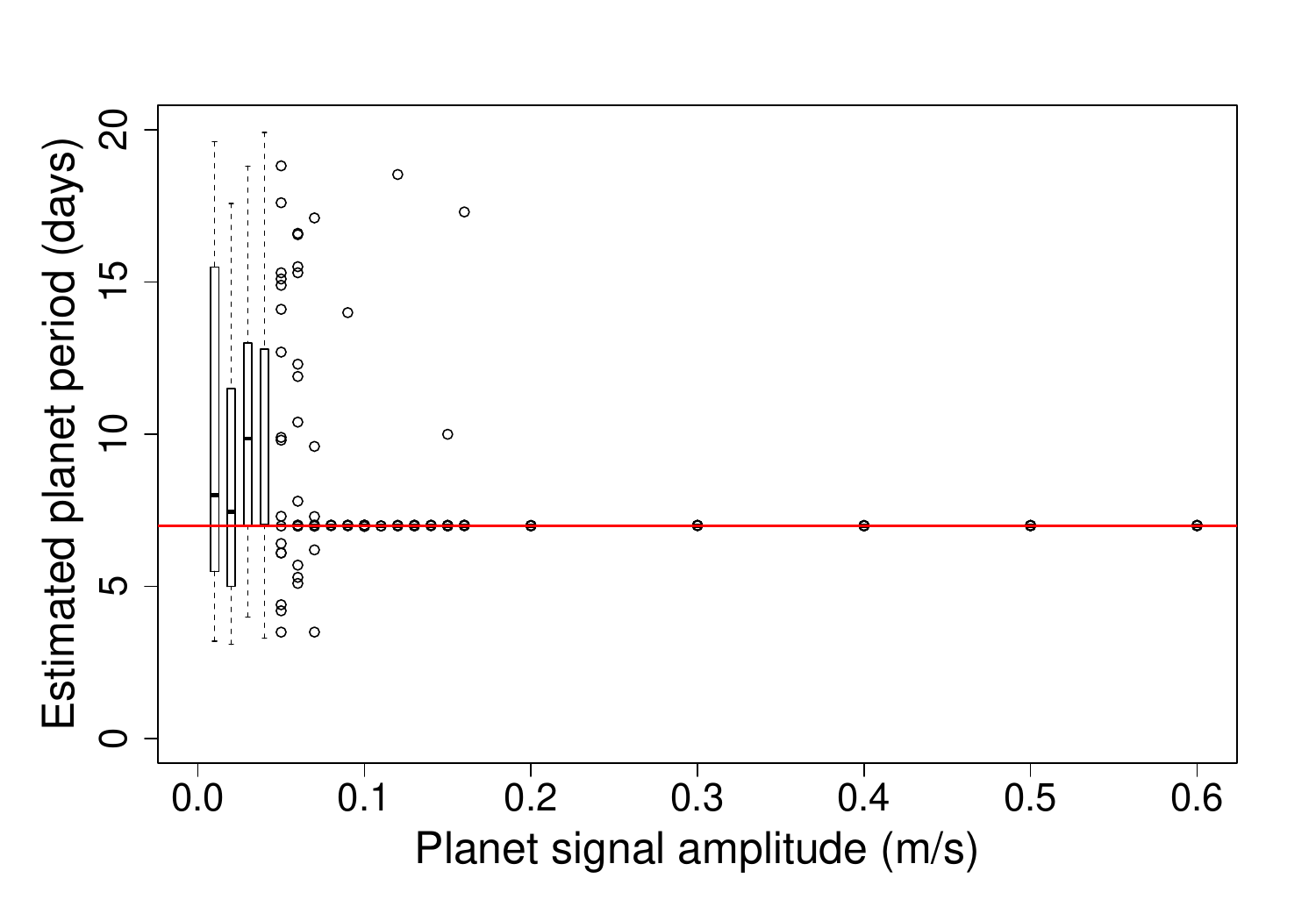}
\end{subfigure}%
\begin{subfigure}{.49\textwidth}
  \centering
  \includegraphics[width=1\linewidth]{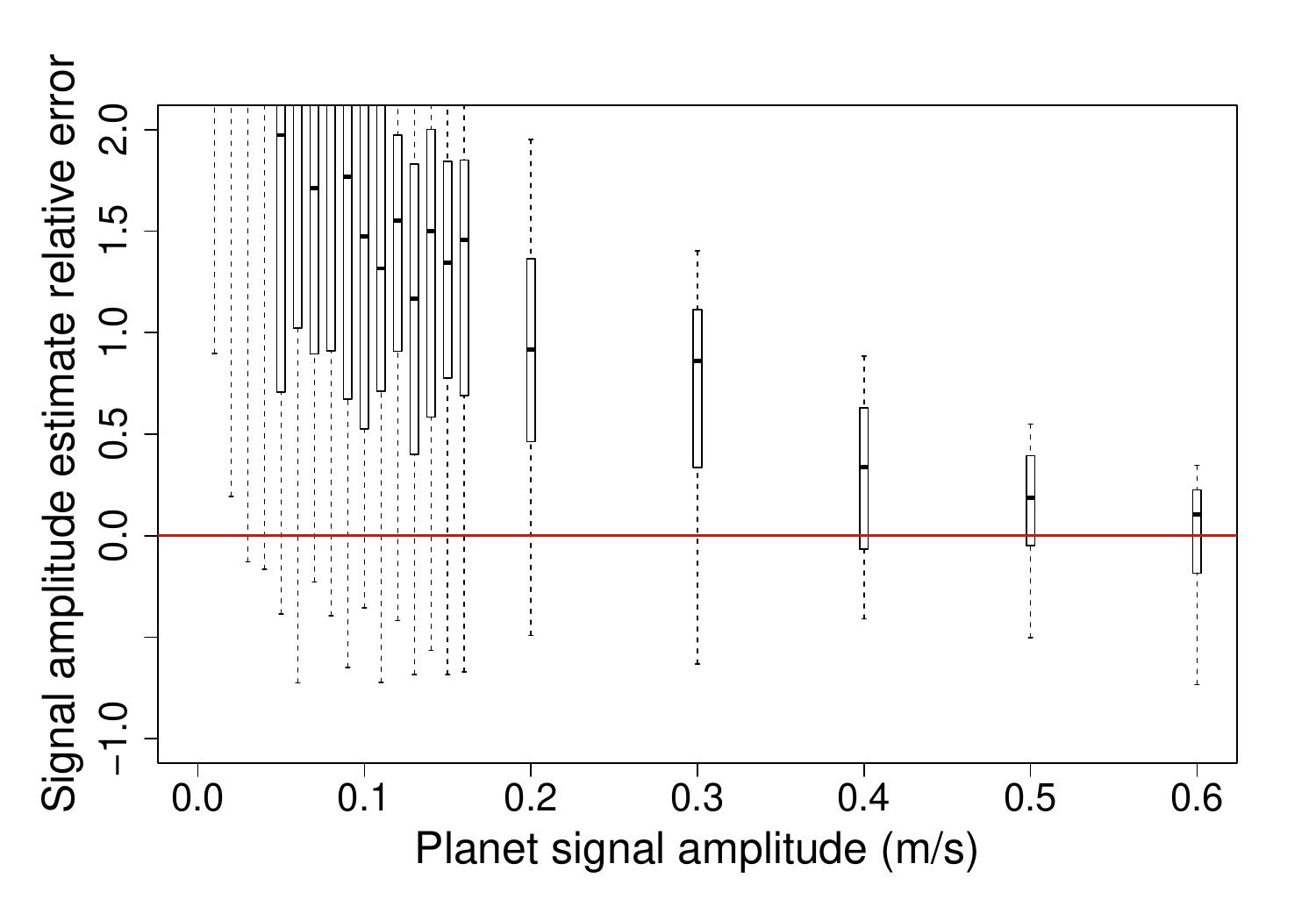}
\end{subfigure}
\caption{The orbital period MLE in days  against the true planet RV amplitude (left); and the relative error of the MLE of the planet RV amplitude $K$ against the true planet RV amplitude (right). In the left panel the red line shows the true orbital period of 7 days, and in the right panel it indicates zero relative error.
\label{fig:estimation}}
\end{figure}

In addition to detecting planets it is of interest to infer their properties. Figure \ref{fig:estimation} summarizes the performance of the MLEs of the orbital period $\tau_p$ and velocity amplitude $K$  under the AIC-1 stellar activity model for the simulations used in Figure \ref{fig:rajpaul_power_tau7}. In Figure \ref{fig:estimation} each boxplot shows the estimates of $\tau_p=7$ (right panel) or the relative errors in estimating $K$ (left panel)  for a given simulation value of $K$. For larger values of $K$, we can see that both $\tau_p$ and $K$  are reasonably well inferred. Since the detection threshold for the AIC-1 model is about $0.065$ m/s, the model may detect some planets even when it is unable to accurately infer their properties. This is to be expected because planet detection requires less information than inferring planet properties. 

\section{Parameters estimates for the R-AIC-2 model}\label{app:rcv3_paras}

In Section \ref{sec:rajpaul_comparison}, we considered modeling the \citetalias{rajpaul2015gaussian} activity indicators, as opposed to our PCA based indicators, and applied Algorithm 3. The R-AIC-2 model was the best model identified and its non-zero coefficients and their maximum likelihood estimates are given in Table \ref{tab:rcv3_fit}. A number of other models were found to yield very similar planet detection power.  

\begin{table}[t]
\caption{Maximum likelihood estimates of the R-AIC-2 stellar activity model coefficients. Blank entries mean the coefficients are set to zero. All the outputs were normalized, but for interpretability the $u(t)$ coefficient estimates in m/s are given in parentheses.\label{tab:rcv3_fit}}
\centering
\begin{tabular}{c|cccc}
& $X$ coeff ($a_{j1}$)& $\dot{X}$ coeff ($a_{j2}$)& $\ddot{X}$ coeff ($a_{j3}$)& $Z_j$ coeff ($a_{j4}$)\\
  \hline
$u(t)$ (m/s) &   & & 0.37 (0.96) &  \\ 
\text{Norm. flux} &0.05 & -0.29 & & 0.14 \\ 
\text{BIS} & & 0.12 & -0.32 & 0.17\\
   \hline
\end{tabular}
\end{table}

Interestingly, the model uses the extra GP components $Z_1$ and $Z_2$. Investigation showed that at least $Z_2$ is  needed in order to properly capture the \citetalias{rajpaul2015gaussian} indicators, which do not have the same level of symmetry as our PCA based indicators (in particular, the BIS minima are not as low as the maxima are high, see the right panel of Figure \ref{fig:rajpaul_power}). Models that only use $X(t)$ and its first two derivatives  fit the \citetalias{rajpaul2015gaussian} indicators comparatively poorly, including the  \citetalias{rajpaul2015gaussian} model (see their Figure 3).

\section{Parameters estimates for the 10k-AIC-1 model}\label{app:10kaic1_paras}

Figure \ref{fig:noisy_power} in Section \ref{sec:generalizing} plots the detection power for the 10k-AIC-1 model, in the case where all null and test datasets have a 10k MSH spot. This model was obtained by again running our model selection procedure, Algorithm 3, but where  all the datasets had a 10k MSH spot. The MLEs  of the model coefficients are given in Table \ref{tab:10kaic1_fit}.

\begin{table}[t]
\caption{Maximum likelihood estimates of the 10k-AIC-1 stellar activity model coefficients. Blank entries mean the coefficients are set to zero. All the outputs were normalized, but for interpretability the $u(t)$ coefficient estimates in m/s are given in parentheses.\label{tab:10kaic1_fit}}
\centering
\begin{tabular}{c|cccc}
& $X$ coeff ($a_{j1}$)& $\dot{X}$ coeff ($a_{j2}$)& $\ddot{X}$ coeff ($a_{j3}$)& $Z_j$ coeff ($a_{j4}$)\\
  \hline
$u(t)$ (m/s) &   &-0.03 (-0.46) & -0.37 (-5.70) &  \\ 
$q_{\text{PC}1}(t)$ & &  0.33 & &-0.15 \\ 
$q_{\text{PC}1}(t)$ & 0.01 & &
 0.38 & \\
   \hline
\end{tabular}
\end{table}

\end{appendices}

\end{document}